\documentclass[twocolumn,floatfix,showpacs]{revtex4}
\usepackage{eurosym}
\usepackage{graphicx}
\usepackage{amsmath}
\usepackage{amsfonts}
\usepackage{amssymb}
\usepackage[caption=false]{subfig}
\usepackage{float}

\begin{document}

\title{One- and two-dimensional solitons supported by singular modulation of
quadratic nonlinearity }
\author{Vitaly Lutsky and Boris A. Malomed}
\affiliation{Department of Physical Electronics, School of Electrical Engineering,\\
Tel Aviv University, Tel Aviv 69978, Israel}

\begin{abstract}
We introduce a model of one- and two-dimensional (1D and 2D) optical media
with the $\chi ^{(2)}$ nonlinearity whose local strength is subject to
cusp-shaped spatial modulation, $\chi ^{(2)}\sim r^{-\alpha }$, with $\alpha
>0$, which can be induced by spatially nonuniform poling. Using analytical
and numerical methods, we demonstrate that this setting supports 1D and 2D
fundamental solitons, at $\alpha <1$ and $\alpha <2$, respectively. The 1D
solitons have a small instability region, while the 2D solitons have a
stability region at $\alpha <0.5$ and are unstable at $\alpha >0.5$. 2D
solitary vortices are found too. They are unstable, splitting into a set of
fragments, which eventually merge into a single fundamental soliton pinned
to the cusp. Spontaneous symmetry breaking of solitons is studied in the 1D
system with a symmetric pair of the cusp-modulation peaks.
\end{abstract}

\maketitle

\pagenumbering{Roman}

\clearpage

\clearpage

\section{\textbf{Introduction and the model}}

\label{sec:Introduction}

\pagenumbering{arabic}

It is commonly known that external potentials greatly expand the variety of
stable localized objects (solitons and solitary vortices) which may be
created by self-focusing nonlinearities \cite{Yang}. Recently, much
attention has also been attracted to the creation of effective nonlinear
potentials (alias ``pseudopotentials", as they are called in
solid-state physics \cite{pseudo}) by means of spatial modulation of local
strength of the nonlinearity \cite{review}. Most works on this topic have
been dealing with the cubic, alias $\chi ^{(3)}$, nonlinearity. In optical
media, spatially nonuniform Kerr nonlinearity can be induced by an
accordingly designed inhomogeneous density of nonlinearity-inducing dopants
\cite{Kip}, by an inhomogeneous distribution of detuning in a uniform
resonant-dopant density \cite{review}, or in composite media assembled of
different materials \cite{Kominis}. Similar nonlinearity landscapes are
relevant in models of Bose-Einstein condensates, where virtually any
landscape, controlled by the optical Feshbach resonance \cite{Feshbach}, can
be ``painted" in space by a rapidly moving laser beam \cite%
{painting}. In the same context, the nonlinearity can be patterned by means
of the magnetic Feshbach resonance, using magnetic lattices \cite{magnetic}.
The limit case of the spatially inhomogeneous $\chi ^{(3)}$ self-interaction
in the one-dimensional (1D) geometry corresponds to a very narrow strip with
strong nonlinearity, which is embedded into a linear host medium. In that
case, the strongly concentrated nonlinearity may be described by the
delta-function \cite{Azbel,Dong,Nir}. Another variety of the strongly
localized self-focusing $\chi ^{(3)}$ nonlinearity is a discrete linear
lattice with one \cite{Tsironis} or two nonlinear \cite{Valery} sites
embedded into it, or externally coupled to the lattice \cite{Sadreev}.

A more realistic model of the singular modulation of the cubic nonlinearity
in the 1D setting was introduced in Ref. \cite{Barcelona}, with the local
strength featuring a cusp, $\chi ^{(3)}\sim |x|^{-\alpha }$, around the
singular point, $x=0$. This model makes it possible to extend the study of
the onset of collapse in nonlinear wave systems, starting from the theory
developed for the uniform space \cite{VK}. It was also found, by means of
analytical and numerical methods, that the nonlinear Schr\"{o}dinger
equation for wave field $u(x,z)$, with the corresponding self-attractive
cubic term $|x|^{-\alpha }|u|^{2}u$, gives rise to stable 1D solitons in the
range of $0\leq \alpha <1$. In the limit of $\alpha =1$, the solitons
disappear, as their amplitude vanishes, and solitons do not exists at $%
\alpha >1$, when the singularity of the nonlinearity modulation is too
strong. Furthermore, in Ref. \cite{Barcelona} it was demonstrated that the
same singular modulation with $\alpha <1$ allows one to emulate the action
of attractive nonlinearities in \textit{sub-1D} spaces, with effective
dimension $D=2\left( 1-\alpha \right) /\left( 2-\alpha \right) <1$.

As mention in Ref. \cite{Barcelona} too, a natural extension of the analysis
should be the consideration of the singular modulation of the quadratic ($%
\chi ^{(2)}$) nonlinearity in a second-harmonic-generating medium. In
particular, an essential advantage of the consideration of the $\chi ^{(2)}$
nonlinearity is the possibility to extend it to the 2D space, where any
self-focusing cubic nonlinearity would immediately give rise to the
collapse. Another advantage of the consideration of $\chi ^{(2)}$ media is
the fact that the well-elaborated poling technique \cite{poling}-\cite%
{Quadratic_solitons} makes it possible to realize various spatially
modulated profiles of the local strength of the $\chi ^{(2)}$ interactions
in 1D and 2D geometries. The technique of quasi-phase-matching \cite{QPM}
may also help to achieve this purpose, by creating spatial patterns of the
phase mismatch \cite{QPM-landscape}. Such patterns do not directly affect
the local $\chi ^{(2)}$ strength, but they determine the effective local
nonlinearity in terms of the cascading limit \cite%
{Quadratic_solitons_Malomed,Quadratic_solitons}, see Eq. (\ref{cascading})
below.

Self-trapped solitary modes, pinned to a spot carrying strongly localized $%
\chi ^{(2)}$ nonlinearity, which may be approximated by a delta-function, $%
\chi ^{(2)}(x)\sim \delta (x)$ \cite{Canberra}, and modes pinned to two such
spots \cite{Asia}, embedded into the linear medium, were studied previously.
Another variety of systems with the strongly localized $\chi ^{(2)}$
nonlinearity is represented by a linear lattice with one or two sites
carrying the nonlinearity \cite{Valery-chi2}. It is also relevant to mention
a recent analysis of a dissipative 1D system with uniform $\chi ^{(2)}$
nonlinearity and a localized gain region (``hot spot" \cite%
{hot}), in which stable dissipative solitons are pinned to the
``hot spot" \cite{Moscow} (recent reviews of dissipative
solitons supported by locally applied gain are presented in Refs. \cite%
{HotSpot1} and \cite{HotSpot2}).

In this work, we concentrate on 1D and 2D systems with singular spatial
modulation of the $\chi ^{(2)}$ nonlinearity, which is represented by
coefficients $|x|^{-\alpha }$ and $r^{-\alpha }$, respectively. The model is
introduced in Section II. In Section III, we report analytical and numerical
results for the 1D version of the system. It is shown that solitons exist at
$\alpha <1$, and they vanish in the limit of $\alpha =1$. In fact, this
existence region is \emph{broader} than its counterpart in the $\chi ^{(3)}$
model, related to the present one by the cascading limit, which is $\alpha
<1/2$, see Eq. (\ref{cascading}) below. An essential result is the stability
map for the solitons, which contains a small instability region at negative
values of the $\chi ^{(2)}$-mismatch coefficient. In Section IV the 1D model
is extended to include a symmetric pair of singular-modulation peaks. In
that case, the main result is a symmetry-breaking bifurcation \cite{book},
which transforms spatially symmetric solitons into asymmetric ones. The 2D
system is considered in Section V, where it is demonstrated that fundamental
2D solitons exist at $\alpha <2$ (although they have tiny amplitudes for $%
1<\alpha <2$), and are stable at $\alpha <0.5$, according to the stability
map shown below in Fig. \ref{fig.Stability_2D_fundamental}. On the other
hand, 2D vortex solitons, which are also constructed in Section V, are found
to be completely unstable, although the scenario of their instability
development is different from the one previously known for the uniform
medium. The paper is concluded by Section VI.

\section{The model}

Basic models for self-guided beams in $\chi ^{(2)}$ media are well known and
have been studied in detail, as summarized in reviews \cite%
{Quadratic_solitons_Malomed} and \cite{Quadratic_solitons}. In the present
work, we focus on the two-wave (degenerate, alias Type-I) quadratic
interactions, which are described by the following set of scaled equations
for the complex fundamental-frequency (FF) and second-harmonic (SH)
amplitudes, $u$ and $v$, in the presence of the spatial singular modulation
of the $\chi ^{(2)}$ nonlinearity:

\begin{gather}
iu_{z}+\frac{1}{2}\nabla ^{2}u+{r}^{-\alpha }u^{\ast }v=0,  \label{u_2d} \\
2iv_{z}+\frac{1}{2}\nabla ^{2}v-Qv+\frac{1}{2}{r}^{-\alpha }u^{2}=0,
\label{v_2d}
\end{gather}%
where $z$ is the propagation distance, diffraction operator $(1/2)\nabla
^{2} $ acts on transverse coordinates $\left( x,y\right) $, $r\equiv \sqrt{%
x^{2}+y^{2}}$, the asterisk (as well as symbol $\mathrm{c.c.}$ used below)
stands for the complex conjugate, and real coefficient $Q$ represents the
SH-FF mismatch. Positive exponent $\alpha $ determines the spatial
modulation of the nonlinearity coefficient, the standard system
corresponding to $\alpha =0$. By means of an obvious rescaling, we fix the
mismatch parameter at one of the three values:%
\begin{equation}
Q=0,+1,-1.  \label{Q}
\end{equation}

Stationary solutions to Eqs. (\ref{u_2d}), (\ref{v_2d}) with real
propagation constant $k$ are looked for as
\begin{equation}
\left\{ u(x,y,z),v(x,y,z)\right\} =\left\{ e^{ikz}\varphi (x,y),e^{2ikz}\psi
(x,y)\right\} ,  \label{stationary}
\end{equation}%
with functions $\varphi (x,y)$ and $\psi (x,y)$ (which are complex for
vortex modes) obeying the stationary equations:

\begin{gather}
-k\varphi +\frac{1}{2}\nabla ^{2}\varphi +{r}^{-\alpha }\varphi ^{\ast }\psi
=0,  \label{phi_2d_stationary} \\
-4k\psi +\frac{1}{2}\nabla ^{2}\psi -Q\psi +\frac{1}{2}{r}^{-\alpha }\varphi
^{2}=0.  \label{psi_2d_stationary}
\end{gather}%
In the 1D geometry, which corresponds to the planar, rather than bulk,
waveguide with the $\chi ^{(2)}$ nonlinearity, Eqs. (\ref{u_2d}) and (\ref%
{v_2d}) amount to the following equations:

\begin{gather}
iu_{z}+\frac{1}{2}u_{xx}+{\left\vert x\right\vert }^{-\alpha }u^{\ast }v=0,
\label{u} \\
2iv_{z}+\frac{1}{2}v_{xx}-Qv+\frac{1}{2}{\left\vert x\right\vert }^{-\alpha
}u^{2}=0.  \label{v}
\end{gather}%
Accordingly, the 1D version of stationary equations (\ref{phi_2d_stationary}%
) and (\ref{psi_2d_stationary}) is%
\begin{gather}
-k\varphi +\frac{1}{2}\varphi ^{\prime \prime }+|x|^{-\alpha }\psi \varphi
=0,  \label{varphi} \\
-\left( 4k+Q\right) \psi +\frac{1}{2}\psi ^{\prime \prime }+\frac{1}{2}%
|x|^{-\alpha }\varphi ^{2}=0,  \label{psi}
\end{gather}%
where functions $\varphi (x)$ and $\psi (x)$ are real, with the prime
standing for $d/dx$. The above-mentioned cascading limit corresponds to
neglecting $\psi ^{\prime \prime }$ in Eq. (\ref{psi}), which yields%
\begin{equation}
\psi \approx \frac{|x|^{-\alpha }}{2\left( 4k+Q\right) }\varphi ^{2},
\end{equation}%
the substitution of which in Eq. (\ref{varphi}) leads to the stationary
equation with the effective cubic nonlinearity:%
\begin{equation}
-k\varphi +\frac{1}{2}\varphi ^{\prime \prime }+\frac{|x|^{-2\alpha }}{%
2\left( 4k+Q\right) }\left\vert \varphi \right\vert ^{2}\varphi =0.
\label{cascading}
\end{equation}

An obvious corollary of Eqs. (\ref{phi_2d_stationary}), (\ref%
{psi_2d_stationary}) and (\ref{varphi}), (\ref{psi}) is that both 2D and 1D
exponentially localized solutions (solitons) may exist under conditions%
\begin{equation}
k>0,4k+Q>0.  \label{>>}
\end{equation}%
An exception might be provided by \textit{embedded solitons}, for which
solely the former condition, $k>0$, is necessary \cite{emb}. However, the
numerical analysis has not revealed embedded solitons in the present model.

Equations (\ref{u_2d}) and (\ref{v_2d}) conserve the total power (alias
Manley-Rowe invariant) \cite{Quadratic_solitons_Malomed}, \cite%
{Quadratic_solitons}), Hamiltonian, and the total angular momentum:

\begin{widetext}
\begin{eqnarray}
P &=&\int_{-\infty }^{+\infty }\left[ \left\vert u(x)\right\vert
^{2}+4\left\vert v(x)\right\vert ^{2}\right] dx,  \label{P} \\
H &=&\int_{-\infty }^{+\infty }\left\{ \frac{1}{2}\left( \left\vert \nabla
u\right\vert ^{2}+\left\vert \nabla v\right\vert ^{2}\right) +Q\left\vert
v\right\vert ^{2}-\frac{1}{2}r^{-\alpha }\left[ \left( u^{\ast }\right)
^{2}v+u^{2}v^{\ast }\right] \right\} dx.  \label{H} \\
M &=&\frac{i}{2}\int \int \left[ u^{\ast }\left( y\frac{\partial u}{\partial
x}-x\frac{\partial u}{\partial y}\right) +2v^{\ast }\left( y\frac{\partial v%
}{\partial x}-x\frac{\partial v}{\partial y}\right) \right] dxdy+\mathrm{c.c.%
}  \label{M}
\end{eqnarray}%
\end{widetext} Dynamical invariants of the 1D version of the system are
obvious counterparts of expressions (\ref{P}) and (\ref{H}) .

It is relevant to note that, as mentioned in Ref. \cite{Barcelona},
dependence $P(k)$ for stationary solutions with wavenumber $k$ and zero
mismatch, $Q=0$, can be found in a general form on the basis of scaling
properties of Eqs. (\ref{phi_2d_stationary}), (\ref{psi_2d_stationary}) and (%
\ref{varphi}), (\ref{psi}):

\begin{eqnarray}
P_{\mathrm{1D}}^{(Q=0)}\left( k\right) &=&k^{(3/2)-\alpha }P_{\mathrm{1D}%
}^{(Q=0)}\left( k=1\right) ,  \notag \\
~P_{\mathrm{2D}}^{(Q=0)}\left( k\right) &=&k^{1-\alpha }P_{\mathrm{2D}%
}^{(Q=0)}\left( k=1\right) .  \label{exact}
\end{eqnarray}%
However, $Q\neq 0$ breaks the scaling invariance and the validity of Eqs. (%
\ref{exact}), making it necessary to produce $P(k)$ relations in \ fully
numerical form, as is done below.

\section{One-dimensional fundamental solitons}

\subsection{\textbf{The variational approximation (VA)}}

\label{sec:Construction_of_Lagrangian_functional_for_variational_analysis}

We start the consideration with the 1D system, noting that Eqs. (\ref{varphi}%
) and (\ref{psi}) for real functions $\varphi (x)$ and $\psi (x)$ can be
derived from the respective Lagrangian,
\begin{widetext}
\begin{equation}
L=\frac{1}{2}\int_{-\infty }^{+\infty }\left\{ \frac{1}{2}\left[ \left(
\varphi ^{\prime }\right) ^{2}+\left( \psi ^{\prime }\right) ^{2}\right] +%
\left[ k\varphi ^{2}+\left( 4k+Q\right) \psi ^{2}-\left\vert x\right\vert
^{-\alpha }\varphi ^{2}\psi \right] \right\} dx,  \label{L}
\end{equation}%
\end{widetext}
Our first aim is to look for fundamental solitons with the help of the
variational approximation (VA), adopting the Gaussian \textit{ansatz}, with
amplitudes $A,B$ and inverse squared widths $\rho ,\gamma $ of the FF and SH
components \cite{Lisak}:%
\begin{equation}
\varphi (x)=A\exp \left( -\rho x^{2}\right) ,\psi (x)=B\exp \left( -\gamma
x^{2}\right) ,  \label{ansatz}
\end{equation}%
with total power%
\begin{equation}
P=\sqrt{\pi /\left( 2\rho \right) }A^{2}+4\sqrt{\pi /\left( 2\gamma \right) }%
B^{2}.  \label{PVA}
\end{equation}

The substitution of the ansatz into Lagrangian\ (\ref{L}) yields
\begin{widetext}
\begin{equation}
L=\frac{1}{4}\left\{ \sqrt{\frac{2\pi }{\gamma \rho }}\left[
B^{2}(8k+2Q+\gamma )\sqrt{\rho }+A^{2}\sqrt{\gamma }(2k+\rho )\right]
-4\Gamma \left( \frac{1-\alpha }{2}\right) A^{2}B(\gamma +2\rho
)^{(-1+\alpha )/2}\right\} ,  \label{lagrangian_result}
\end{equation}%
\end{widetext}
where $\Gamma $ is the Gamma-function. Two variational equations, $\partial
L/\partial B=\partial L/\partial \left( A^{2}\right) =0$, which follow from
Lagrangian (\ref{lagrangian_result}), produce the following expressions for
the FF and SH amplitudes:

\begin{equation}
A^{2}=\frac{\pi (8k+2Q+\gamma )(2k+\rho )(\gamma +2\rho )^{1-\alpha }}{4%
\sqrt{\gamma \rho }\left[ \Gamma \left( \left( 1-\alpha \right) /2\right) %
\right] ^{2}},  \label{A_sqr}
\end{equation}

\begin{equation}
B=\sqrt{\frac{\pi }{2\rho }}\frac{(2k+\rho )(\gamma +2\rho )^{\left(
1-\alpha \right) /2}}{2\Gamma \left( \left( 1-\alpha \right) /2\right) }.
\label{B}
\end{equation}%
These expressions make sense at $\alpha <1$, predicting that the solitons
disappear at $\alpha =1$, as they yield $A^{2}(\alpha =1)=B(\alpha =1)=0$.
Below, it is shown directly that this happens indeed. It is relevant to
mention that the cascading limit, which amounts to Eq. (\ref{cascading}),
admits the existence of 1D solitons only at $\alpha <1/2$, according to Ref.
\cite{Barcelona}.

The remaining variational equations, $\partial L/\partial \rho =\partial
L/\partial \gamma =0$, lead to a system of coupled quadratic equations for
the inverse squared widths, $\rho $ and $\gamma $, where the amplitudes were
eliminated by means of Eqs. \ref{A_sqr} and (\ref{B}):
\begin{widetext}
\begin{equation}
\begin{cases}
\rho \left[ \gamma +2(2-\alpha )\rho \right] -2k(\gamma +2\alpha \rho )\ =0,
\\
\gamma \left[ \left( 2\alpha -3\right) \gamma -2\rho \right] +8k(\left[
(2\alpha -1)\gamma +2\rho \right] +2Q\left[ \left( 2\alpha -1\right) \gamma
+2\rho \right] =0.%
\end{cases}
\label{var-eqns}
\end{equation}%
\end{widetext}These equations can be readily solved numerically.

\subsection{The form of 1D solitons around\textbf{\ }$x=0$}

The singular modulation of the nonlinearity makes it necessary to analyze
the structure of the solitons solutions at $x\rightarrow 0$, similar to how
it was done for the $\chi ^{(3)}$ model in Ref. \cite{Barcelona}.
Accordingly, the solution is sought for in the form of an expansion,%
\begin{equation}
\varphi (x)=\varphi _{0}-\varphi _{1}|x|^{2-\alpha }+...,~\psi (x)=\psi
_{0}-\psi _{1}|x|^{2-\alpha }+...,  \label{vicinity}
\end{equation}%
where $\alpha <2$ is implied. It is easy to see that this expansion
corresponds to a local \emph{maximum} of both fields at $x=0$ (i.e., to
possible soliton solutions) under the condition that%
\begin{equation}
\varphi _{1}/\varphi _{0}>0,~\psi _{1}/\psi _{0}>0.  \label{>0}
\end{equation}%
A simple calculation, performed for $x\rightarrow 0$, yields the following
results for expansion coefficients $\varphi _{1}$ and $\psi _{1}$:%
\begin{eqnarray}
\frac{\varphi _{1}}{\varphi _{0}} &=&\frac{2\psi _{0}}{\left( 2-\alpha
\right) \left( 1-\alpha \right) },  \label{10} \\
\psi _{1} &=&\frac{\varphi _{0}^{2}}{\left( 2-\alpha \right) \left( 1-\alpha
\right) }.  \label{100}
\end{eqnarray}%
An immediate conclusion following from Eqs. (\ref{10}) and (\ref{100}) is
that condition (\ref{>0}) of having a maximum at $x=0$ amounts to the
following inequalities:
\begin{subequations}
\label{max}
\begin{eqnarray}
\psi _{0} &>&0~~\mathrm{at~}~0<\alpha <1, \\
\psi _{0} &<&0~~\mathrm{at~}~1<\alpha <2.
\end{eqnarray}%
Actually, only in the former case, $0<\alpha <1$, the solitons exist, while
in the latter case, $1<\alpha <2$, the singularity is too strong and cannot
support solitons. This fact can be simply explained by the observation that
the average value of the scaled $\chi ^{(2)}$ nonlinearity coefficient in a
region around $x=0$, $|x|<L$, is determined by integral
\end{subequations}
\begin{equation}
\left\langle \chi ^{(2)}\right\rangle _{\mathrm{1D}}\equiv \frac{1}{2L}%
\int_{-L}^{+L}|x|^{-\alpha }dx=\frac{L^{-\left( 1+\alpha \right) }}{1-\alpha
},  \label{singular}
\end{equation}%
which converges at $\alpha <1$ and diverges at $\alpha \geq 1$. Note that
the VA equations (\ref{A_sqr}) and (\ref{B}) lead to exactly the same
conclusion, showing that the solitons exist solely at $\alpha <1$, even if
the expansion of the variational ansatz (\ref{ansatz}) does not exactly
correspond to exact results represented by Eqs. (\ref{vicinity}) and (\ref%
{10}), (\ref{100}). In principle, a more accurate ansatz may be taken as $%
\varphi (x)=A\exp \left( -\rho x^{2-\alpha }\right) ,\psi (x)=B\exp \left(
-\gamma x^{2-\alpha }\right) ,$ to comply with Eq. (\ref{vicinity}), but the
VA takes quite a cumbersome form in this case.

\subsection{\textbf{Numerical results}}

Using the VA-predicted wave forms as an input, it is straightforward to
construct a family of fundamental-soliton solutions of Eqs. (\ref{varphi})
and (\ref{psi}) by means of the Newton's method. Figure \ref%
{fig:Numerically_found_profile_1d} displays typical examples of the
fundamental 1D solitons, both stable and unstable.

\begin{figure*}[tbp]
\centering
\subfloat[]{
            \includegraphics[width=0.49\textwidth]{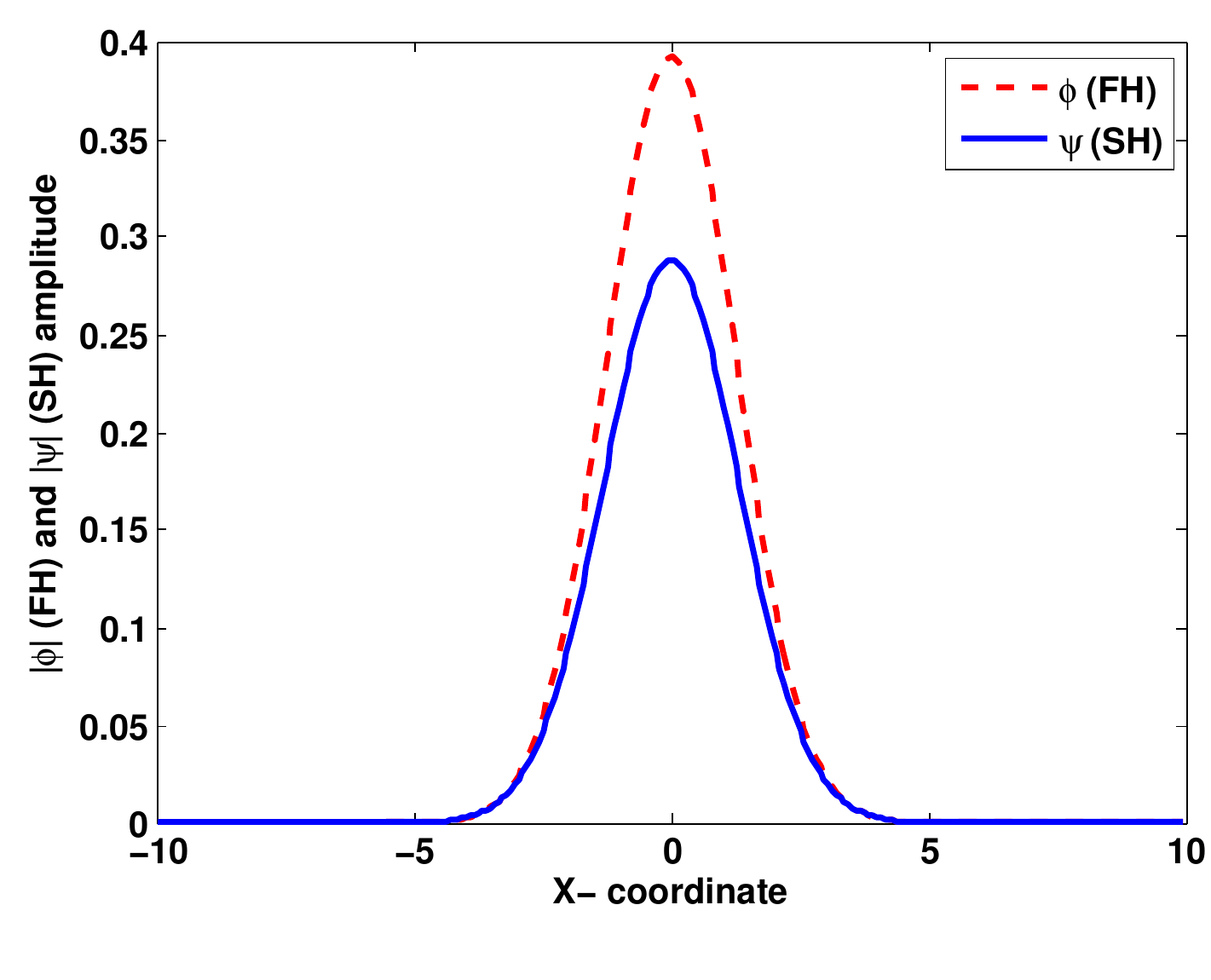}
                        \label{fig:Numerically_found_profile_1d_stable}
}
\subfloat[]{
            \includegraphics[width=0.49\textwidth]{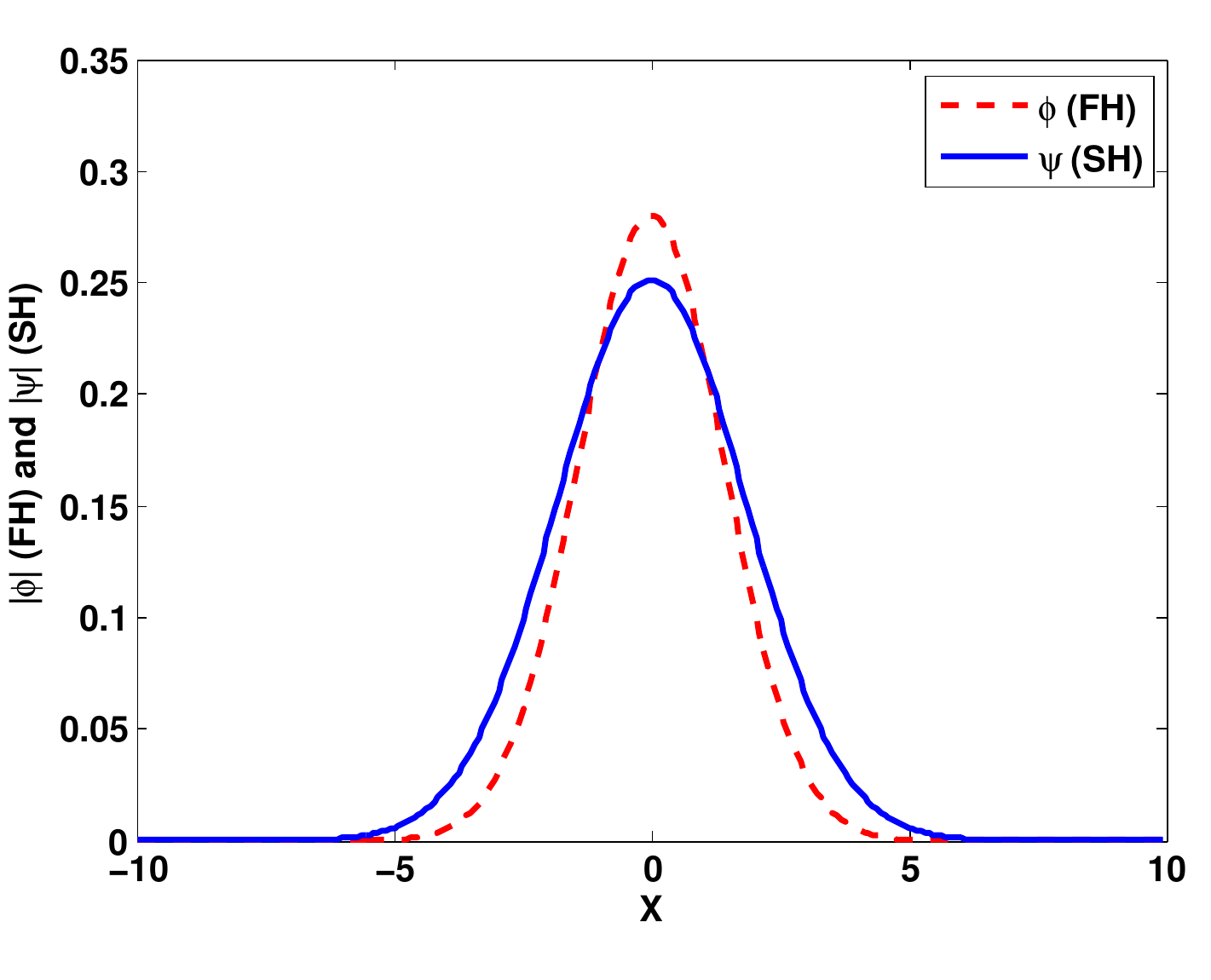}
                        \label{fig:Numerically_found_profile_1d_unstable}
}
\caption{(Color online) (a) Numerically found profiles of the FH and SH
components for a stable 1D soliton, with $Q=-1$, $k=0.32$, $\protect\alpha $
$=0.5$. (b) The same for an unstable 1D soliton, with $Q=-1$, $k=0.28$, $%
\protect\alpha =0.5$.}
\label{fig:Numerically_found_profile_1d}
\end{figure*}

\begin{figure*}[tbp]
\centering
\subfloat[]{
            \includegraphics[width=0.49\textwidth]{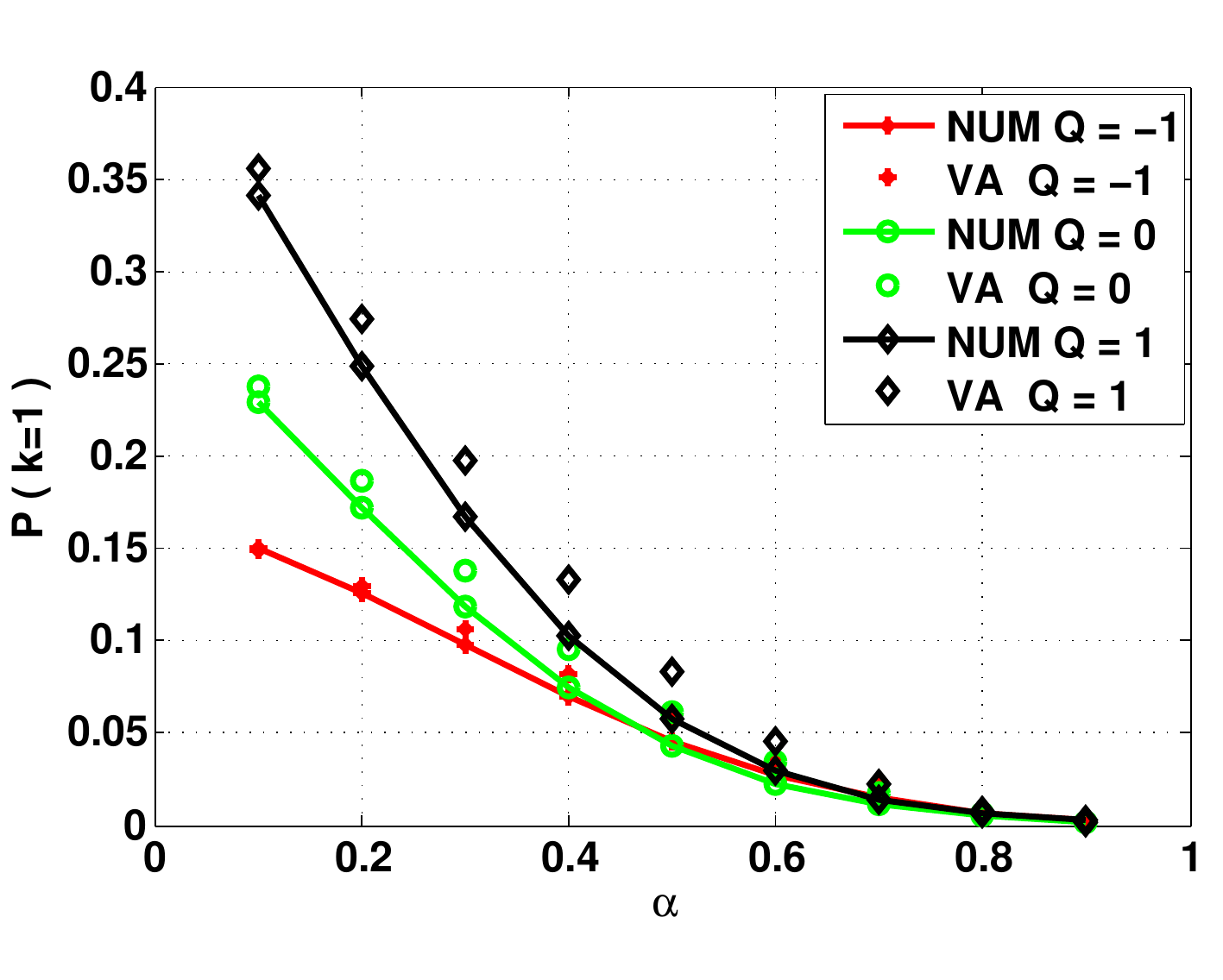}
                        \label{fig:P_vs_alpha_1d}
}
\subfloat[]{
            \includegraphics[width=0.49\textwidth]{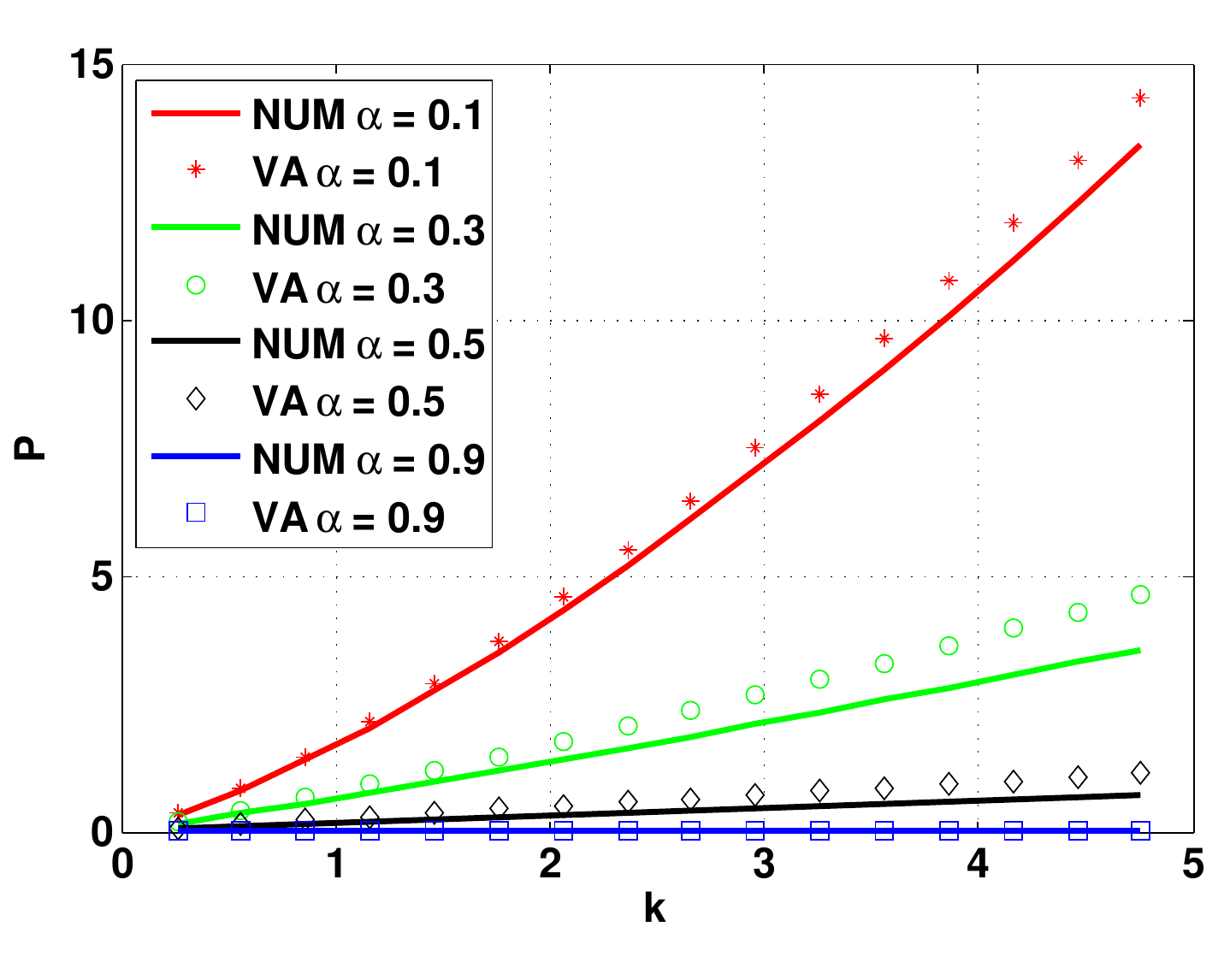}
                        \label{fig:P_vs_k_different_alpha_1d}
}
\caption{(Color online) (a) The total power of 1D fundamental solitons, $P$,
vs. the singularity exponent, $\protect\alpha $, at a fixed propagation
constant, $k=1$, and different values of the mismatch parameter, $Q$. (b) $%
P(k)$ for $Q=+1$ and several fixed values of $\protect\alpha $. Both
numerical results and their counterparts generated by the VA [see Eqs. (%
\protect\ref{ansatz})-(\protect\ref{var-eqns})] are displayed.}
\label{fig:Properties of numerically found solitons 1D}
\end{figure*}

In Fig. \ref{fig:Properties of numerically found solitons 1D}, the family is
characterized by the dependence of the total power, $P$, defined as per Eq. (%
\ref{P}), on the singular-modulation exponent, $\alpha $, and on the
propagation constant, $k$. The results labeled by VA are obtained from the
Eqs. (\ref{PVA}), (\ref{A_sqr}), and (\ref{B}), with $\rho $ and $\gamma $
produced by a numerical solution of Eq. (\ref{var-eqns}). Figure \ref%
{fig:Properties of numerically found solitons 1D} demonstrates that the VA
provides a reasonable, although imperfect, accuracy, in comparison with the
numerical findings.

In accordance with what was said above, Fig. \ref{fig:Properties of
numerically found solitons 1D}(a) confirms that solitons indeed vanish at $%
\alpha =1$ and do not exist at $\alpha >1$. The same figure demonstrates
that dependence of the family on mismatch $Q$ [see Eq. (\ref{Q})] is
relatively weak. However, there is an essential difference between $Q=-1$
and $Q=0,+1$, as concerns stability of the solitons, see below.{\Large \ }

\begin{figure*}[tbp]
\centering%
\subfloat[]{
            \includegraphics[width=0.49\textwidth]{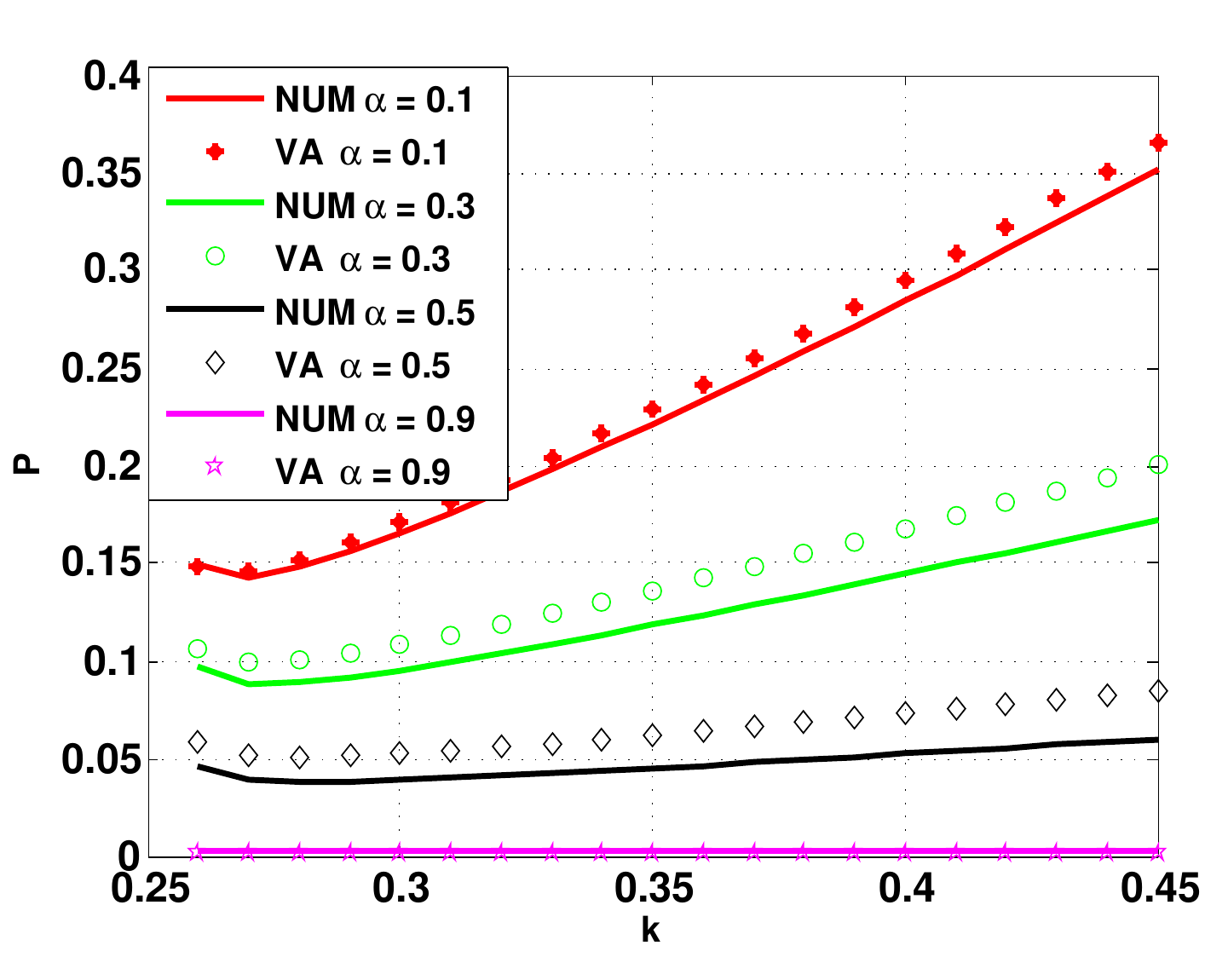}
 }
\subfloat[]{
            \includegraphics[width=0.49\textwidth]{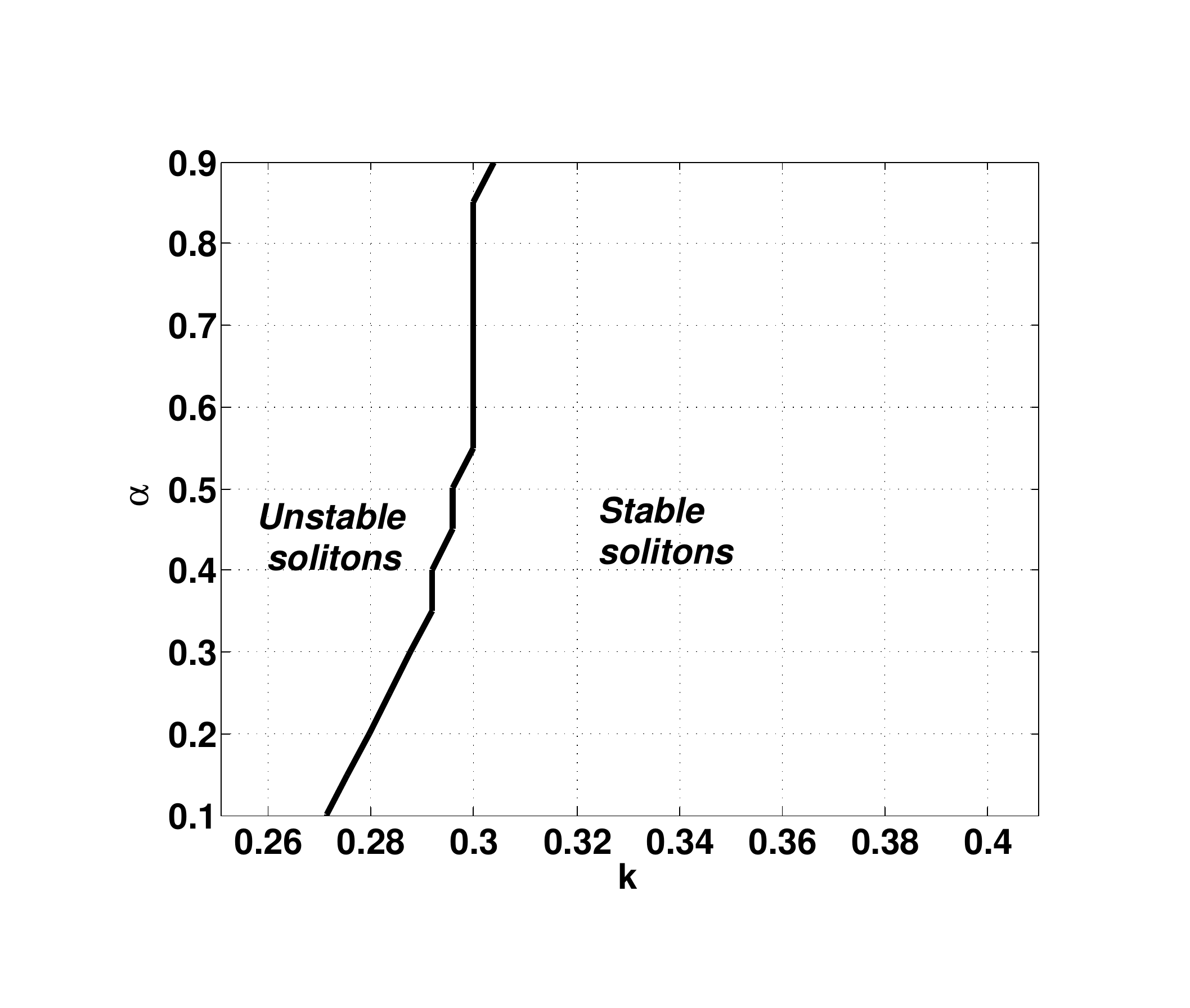}
}
\caption{(a) $P(k)$ for the fundamental 1D solitons with the negative
mismatch, $Q=-1$, and several fixed values of the singular-modulation
exponent, $\protect\alpha $. This plot demonstrates that there is a region
with $dP/dk<0$, where the VK stability criterion does not hold. (b) The
stability map for the solitons with $Q=-1$ in the plane of $k$ and $\protect%
\alpha $ (for $Q=0$ and $Q=+1$, all the fundamental solitons are stable).
The instability boundary is produced in the same form by the VK criterion
and by the computation of stability eigenvalues for modes of small
perturbations.}
\label{fig:Stability_map}
\end{figure*}

An obvious feature observed in Fig. \ref{fig:Properties of numerically found
solitons 1D}(b) (for $Q=+1$) is the positive slope of the curves, i.e., $%
dP/dk>0$, hence the soliton family satisfies the Vakhitov-Kolokolov (VK)
criterion, which is a necessary condition for their stability \cite{VK}.%
\textbf{\ }The same conclusion pertains to the solitons found at $Q=0$,
which is actually an exact result, according to Eq. (\ref{exact}). On the
other hand, plots $P(k)$ for $Q=-1$, shown in Fig. \ref{fig:Stability_map}%
(b), exhibit a region where the VK criterion is not satisfied.

To accurately check the stability of the 1D fundamental solitons, we have
computed instability growth rates for eigenmodes of small perturbations
added to the stationary solitons (the procedure is described in Appendix \ref%
{App:AppendixA}.) This analysis has confirmed that the VK criterion is not
only necessary but, as a matter of fact, also sufficient for the stability
of the 1D fundamental solitons in the present context. Thus, the solitons
are completely stable for $Q=+1$ and $Q=0$, while the boundary between the
stable and unstable solitons for $Q=-1$ is shown, in the plane of $\left(
k,\alpha \right) $, in Fig. \ref{fig:Stability_map}(b). In fact, the narrow
instability region at $\alpha =0$ is akin to the known narrow instability
domain for fundamental 1D solitons in the standard $\chi ^{(2)}$ system \cite%
{Quadratic_solitons_Malomed,Quadratic_solitons}.

\begin{figure*}[tbp]
\centering
\subfloat[]{
         \includegraphics[width=0.49\textwidth]{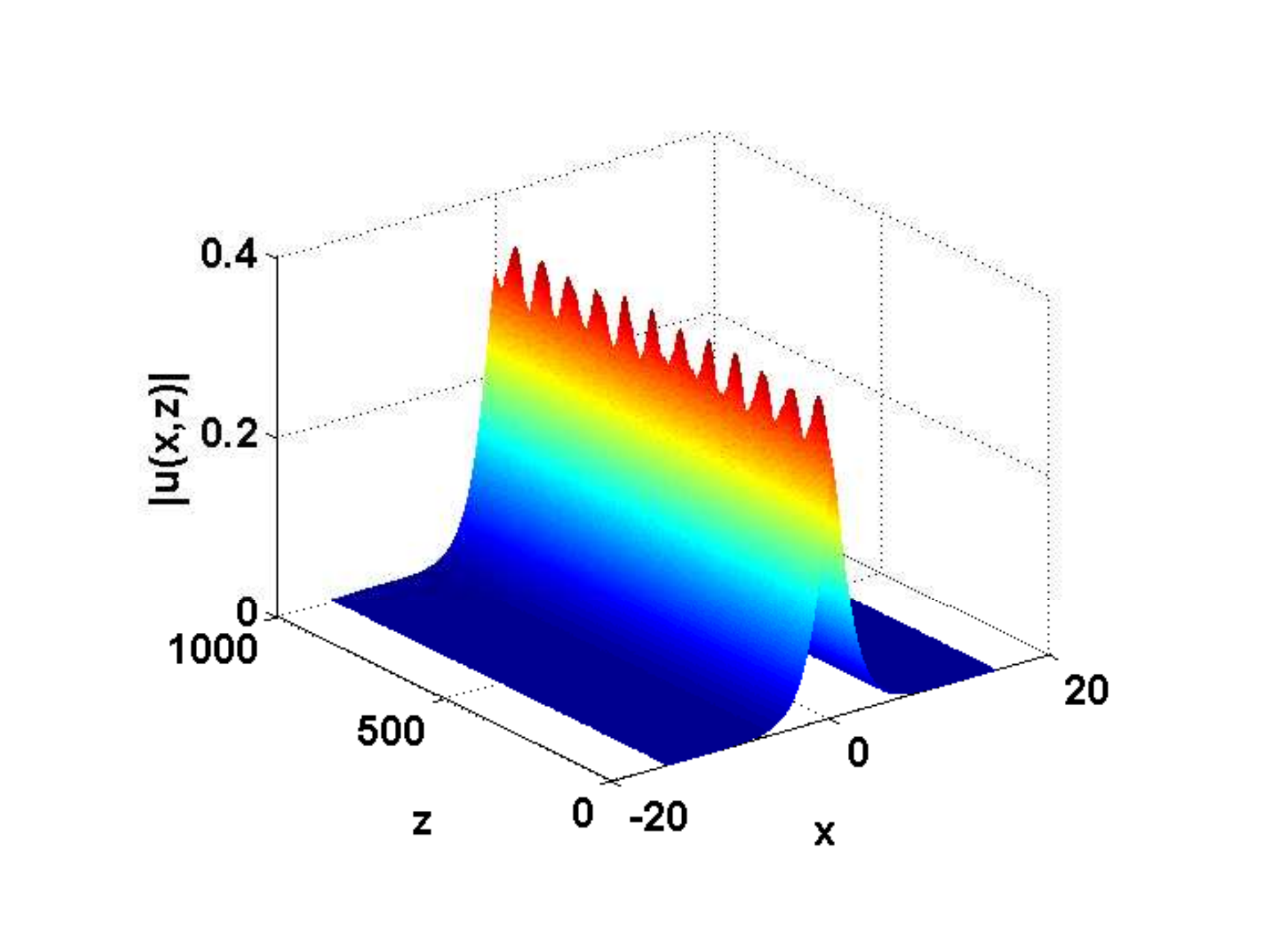}
         \label{fig:Direct_unstable_1_first_harmonic_alpha_0_5_k_0_28_plus_pert}
}
\subfloat[]{
         \includegraphics[width=0.49\textwidth]{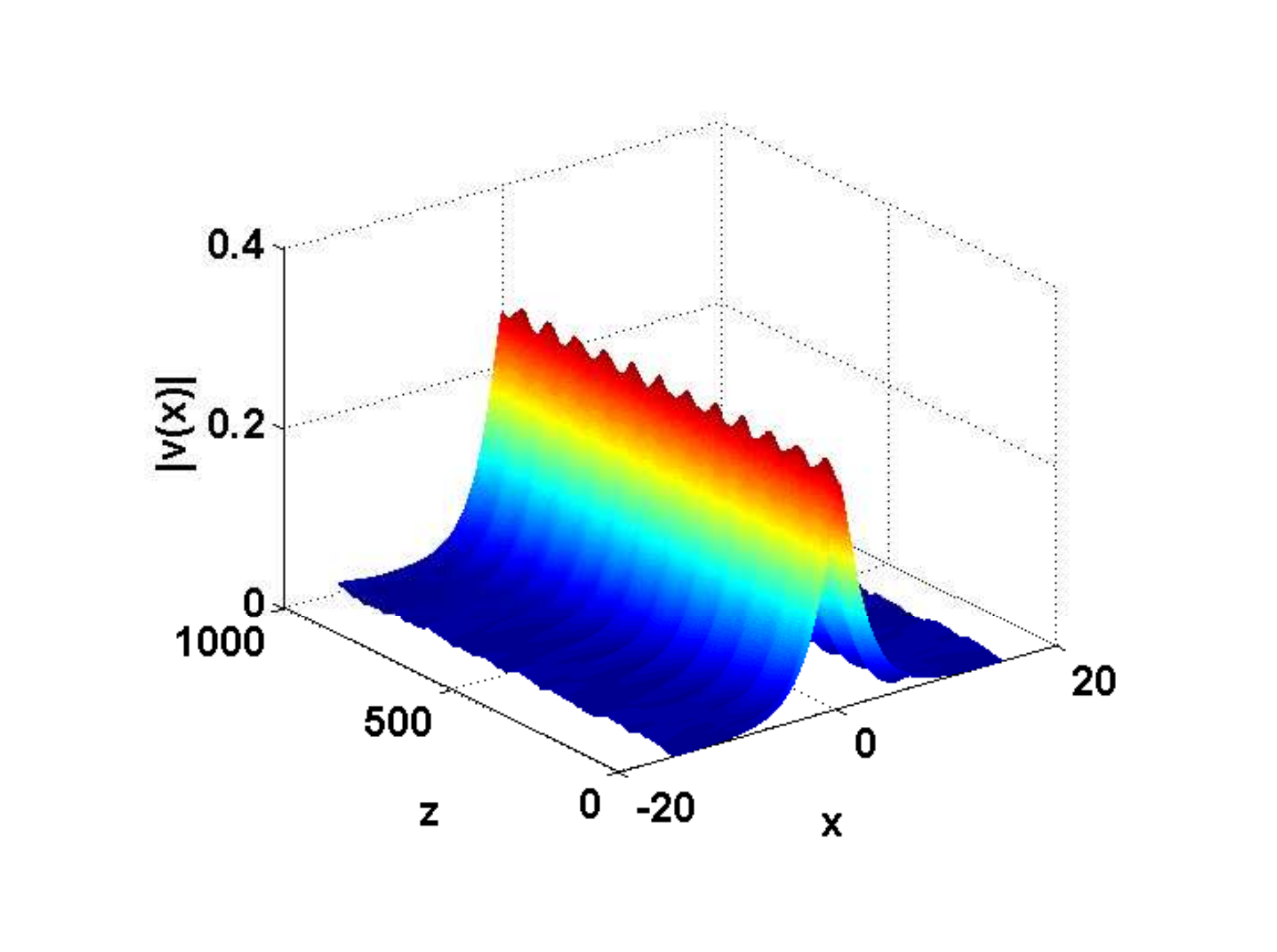}
         \label{fig:Direct_unstable_1_second_harmonic_alpha_0_5_k_0_28_plus_pert}
}
\caption{(Color online) A typical example of the evolution of the FF (a) and
SH (b) components of an unstable fundamental soliton, at $k=0.28$, $Q=-1$
and $\protect\alpha =0.5$, whose integral power exceeds that of a
co-existing stable soliton $(P>0.04)$. In this case, the evolution leads to
the formation of a breather close to the stable soliton.}
\label{fig:Unstable_1d_examples}
\end{figure*}

The predicted stability and instability of the 1D fundamental solitons was
also verified by direct simulations. It has been found that, if the integral
power (\ref{P}) of unstable solitons is smaller than the power of their
stable counterparts, the unstable solitons suffer complete decay (not shown
here in detail). On the other hand, Fig. \ref{fig:Unstable_1d_examples}
demonstrates that unstable solitons, whose integral power exceeds that for
coexisting stable solitons, do not decay, but rather transform into
breathers oscillating around stable solitons.

\section{The 1D \textbf{model with a pair of singular-modulation peaks }}

To introduce the model with the double-peak spatial modulation, we replace
the underlying equations (\ref{u}), (\ref{v}) by more general ones,%
\begin{gather}
iu_{z}+\frac{1}{2}u_{xx}+g(x)u^{\ast }v=0,  \label{u_gen_g} \\
2iv_{z}+\frac{1}{2}v_{xx}-Qv+\frac{1}{2}g(x)u^{2}=0,  \label{v_gen_g}
\end{gather}
and select the double-peak modulation function $g(x)$ in the form of

\begin{equation}
g(x)=\left\vert x-\Delta \right\vert ^{-\alpha }+\left\vert x+\Delta
\right\vert ^{-\alpha },  \label{General_modulation_g}
\end{equation}%
where the separation between the peaks is $2\Delta $, while the mismatch
parameter may be scaled, as above, to one of the three fixed values (\ref{Q}%
).

\begin{figure*}[tbp]
\centering
\subfloat[]{
            \includegraphics[width=0.33\textwidth]{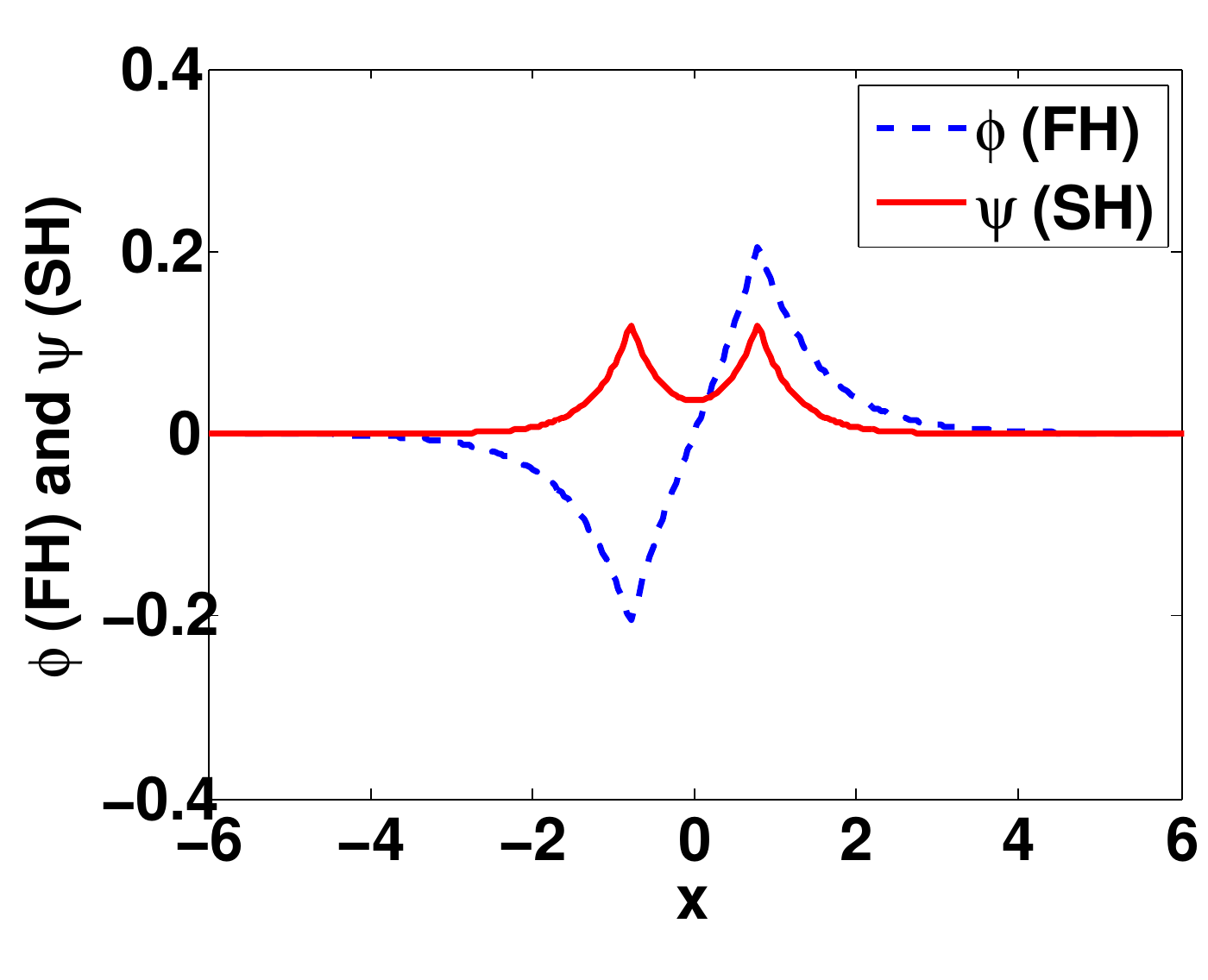}
}
\subfloat[]{
            \includegraphics[width=0.33\textwidth]{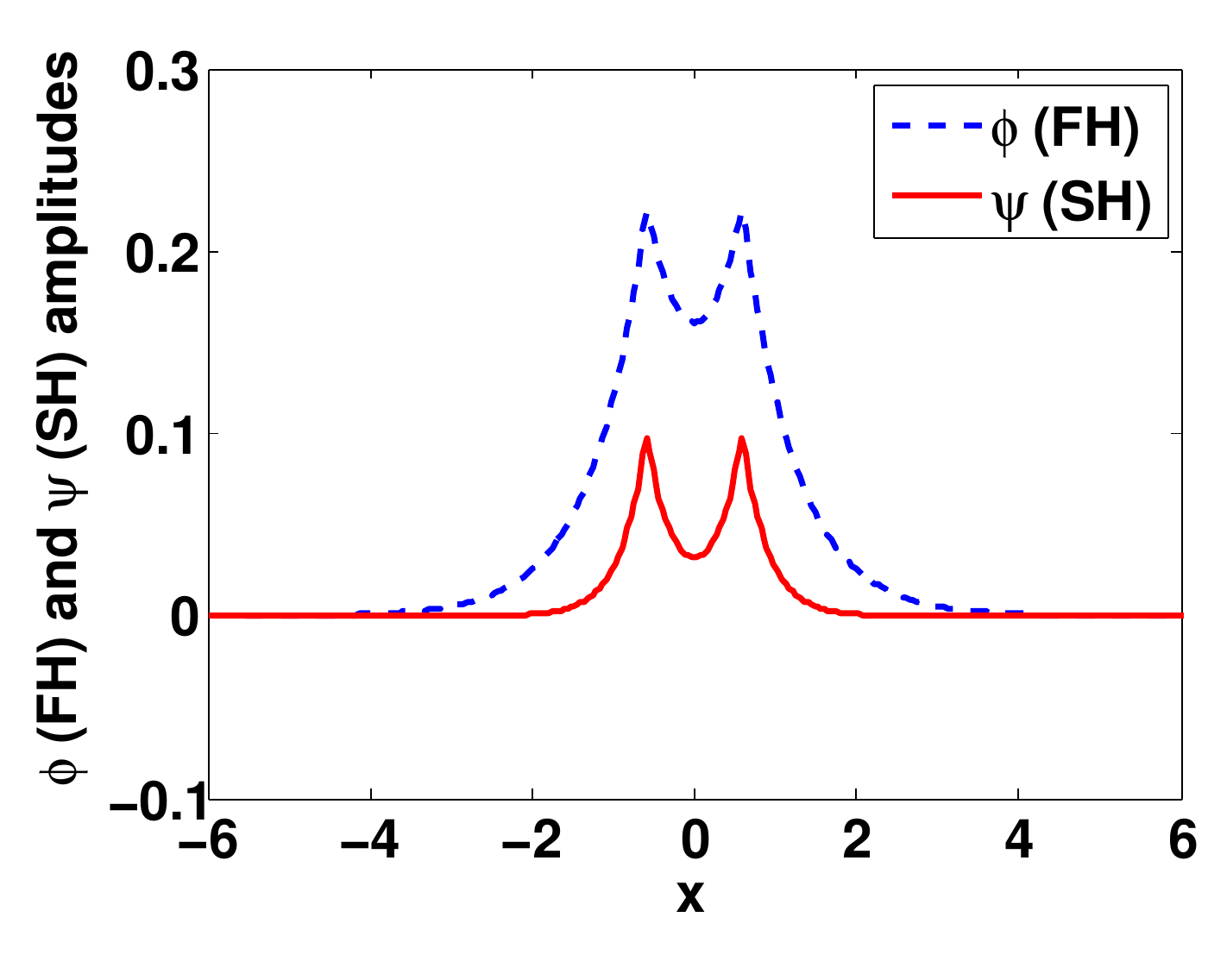}
}
\subfloat[]{
            \includegraphics[width=0.33\textwidth]{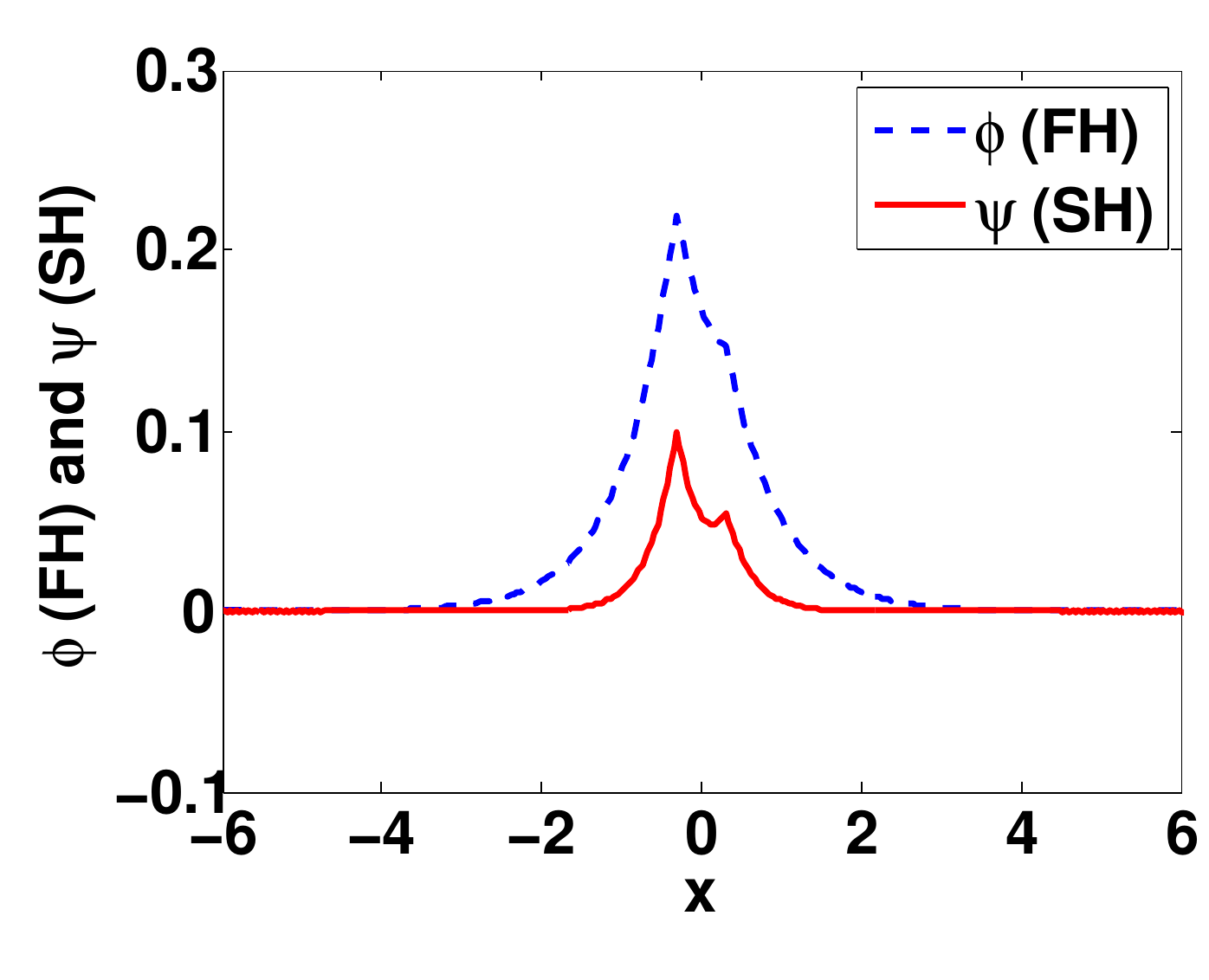}
}
\caption{(Color online) Typical profiles of the FF\ and SH components of
stable 1D solitons supported by the double-peak spatial modulation of the
nonlinearity: (a) a twisted soliton ($\Delta =1.17$); (b) a symmetric
soliton ($\Delta =0.59$); (c) a strongly asymmetric soliton ($\Delta =0.31$%
). In all the cases, $Q=+1$, $\protect\alpha =0.5$, and $k=1.2.$ $\Delta $
is the respective value of the half-separation between the peaks.}
\label{fig:Soliton_Shapes}
\end{figure*}

Numerical solution of the stationary version of Eqs. (\ref{u_gen_g}) and (%
\ref{v_gen_g}) reveals three types of stationary soliton solutions, namely,
symmetric ones with maxima separated by distance $2\Delta $, \textit{%
asymmetric} solitons, for which the amplitudes are different at $x=+\Delta $
and $-\Delta $, and \textit{twisted} solitons, in which the FF component is
antisymmetric, with opposite amplitudes at $x=\pm \Delta $, while the SH
component is symmetric. Typical examples of profiles of all the three types
of the solitons are displayed in Fig. \ref{fig:Soliton_Shapes}.

\begin{figure*}[tbp]
\centering
\subfloat[]{
            \includegraphics[width=0.49\textwidth]{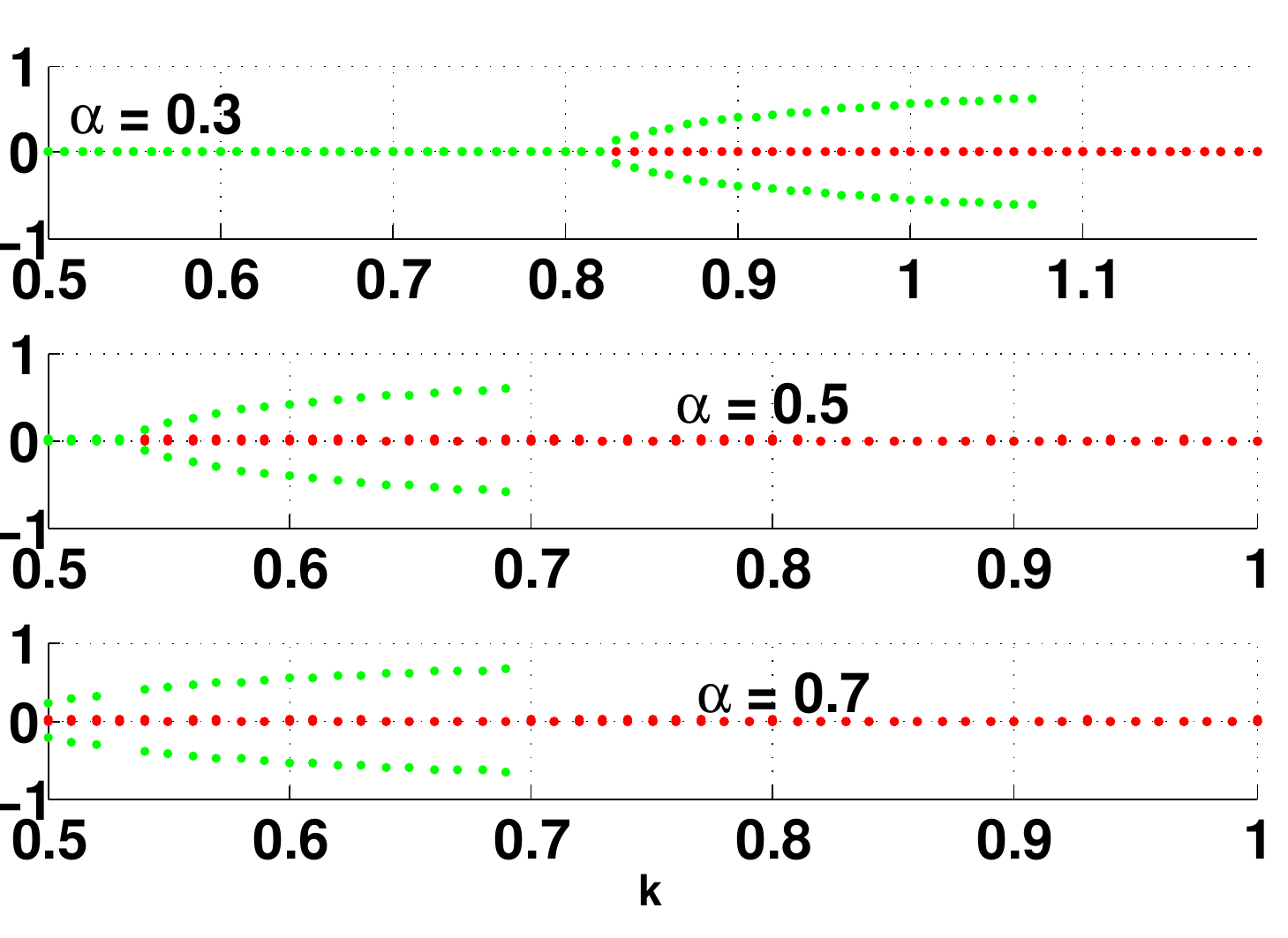}
}
\subfloat[]{
            \includegraphics[width=0.49\textwidth]{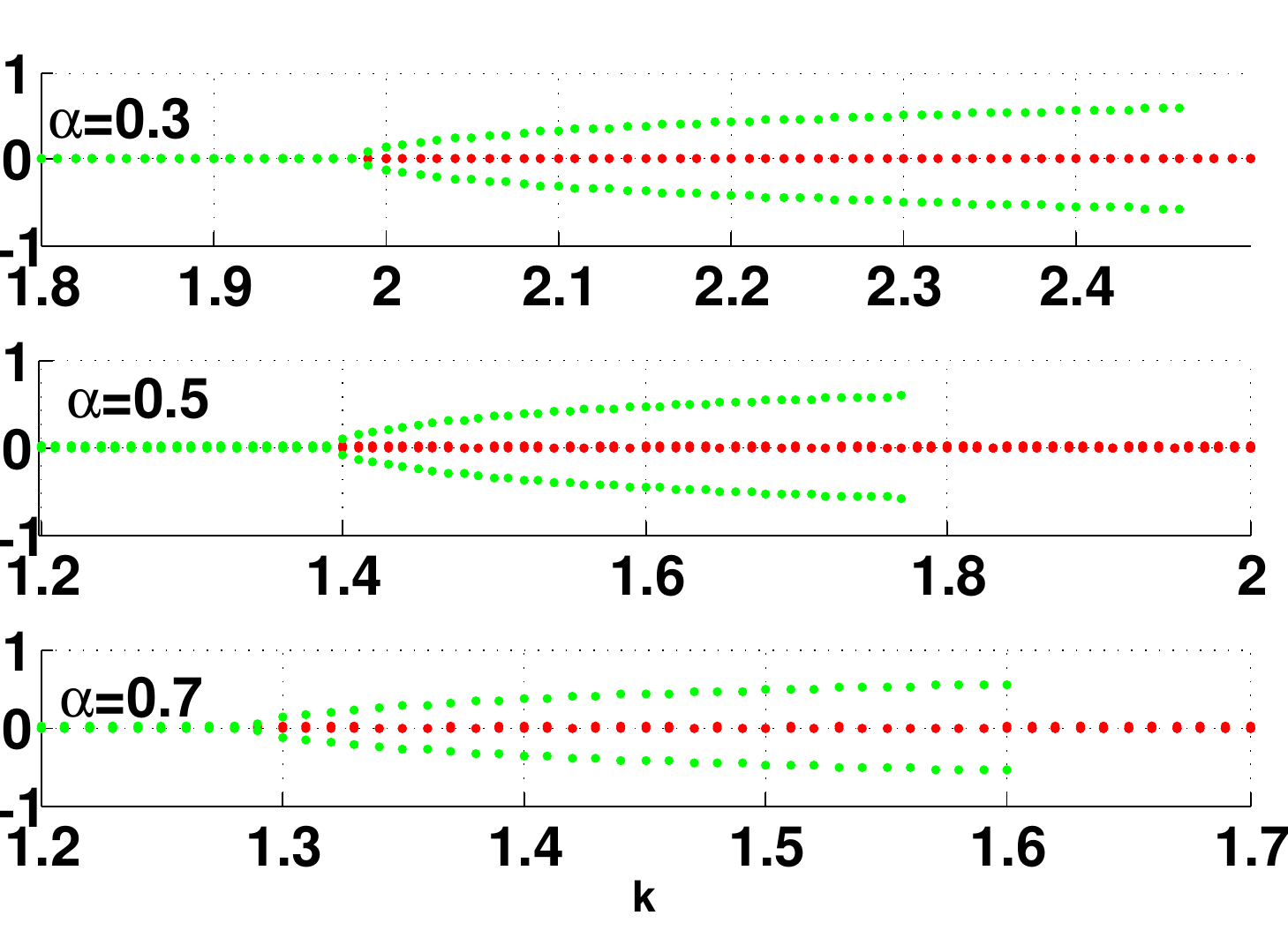}
}
\caption{(Color online) Bifurcation diagrams for the spontaneous symmetry
breaking of symmetric solitons supported by the double-peak spatial
modulation of the nonlinearity, with separation $2\Delta $ between the peaks
and $Q=+1$, at fixed values of the singularity exponent $\protect\alpha $:
(a) $\Delta =0.43$; (b) $\Delta =0.27$. Asymmetry parameter (\protect\ref%
{Theta}) is shown vs. wavenumber $k$ of the solitons. Green and red lines
depict stable and unstable soliton families, respectively (see the text).}
\label{fig:Bifurcations_Q_1}
\end{figure*}

Similar to other models with two separated symmetric peaks of the local $%
\chi ^{(2)}$ \cite{Valery-chi2} and $\chi ^{(3)}$ \cite{Dong,Yasha,Valery}
nonlinearity strength , the asymmetric solitons are generated from the
symmetric ones by the \textit{spontaneous-symmetry-breaking bifurcation}
\cite{book}, which occurs with the increase of separation $2\Delta $ between
the peaks, if other parameters are fixed. The bifurcation is illustrated by
Fig. \ref{fig:Bifurcations_Q_1}, which shows the measure of the asymmetry
between the local powers at the two peaks,
\begin{widetext}
\begin{equation}
\Theta \equiv \frac{\left[ |u(x=\Delta )|^{2}+4|v(x=\Delta )|^{2}\right] -%
\left[ |u(x=-\Delta )|^{2}+4|v(x=-\Delta )|^{2}\right] }{\left[ |u(x=\Delta
)|^{2}+4|v(x=\Delta )|^{2}\right] +\left[ |u(x=-\Delta )|^{2}+4|v(x=-\Delta
)|^{2}\right] },  \label{Theta}
\end{equation}%
\end{widetext}as a function of the soliton's wavenumber, for $Q=+1$ (for $%
Q=0 $ and $-1$ the bifurcation diagrams are quite similar). Coordinates of
the bifurcation points are summarized in Fig. \ref%
{fig:Bifurcation_Split_point}. It is observed that the location of the
bifurcation only weakly depends on mismatch $Q$.

\begin{figure*}[tbp]
\centering
\subfloat[]{
            \includegraphics[width=0.49\textwidth]{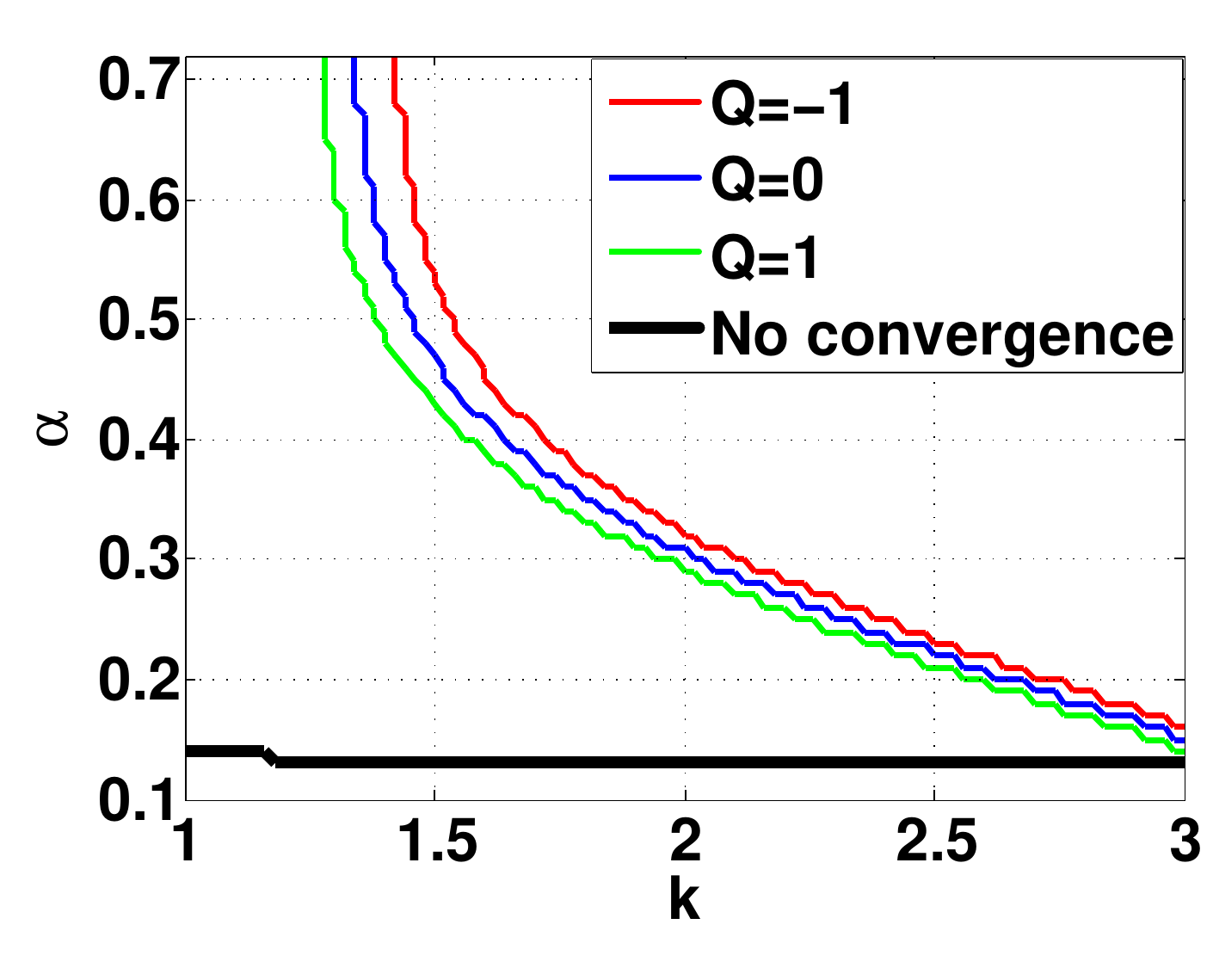}
}
\subfloat[]{
            \includegraphics[width=0.49\textwidth]{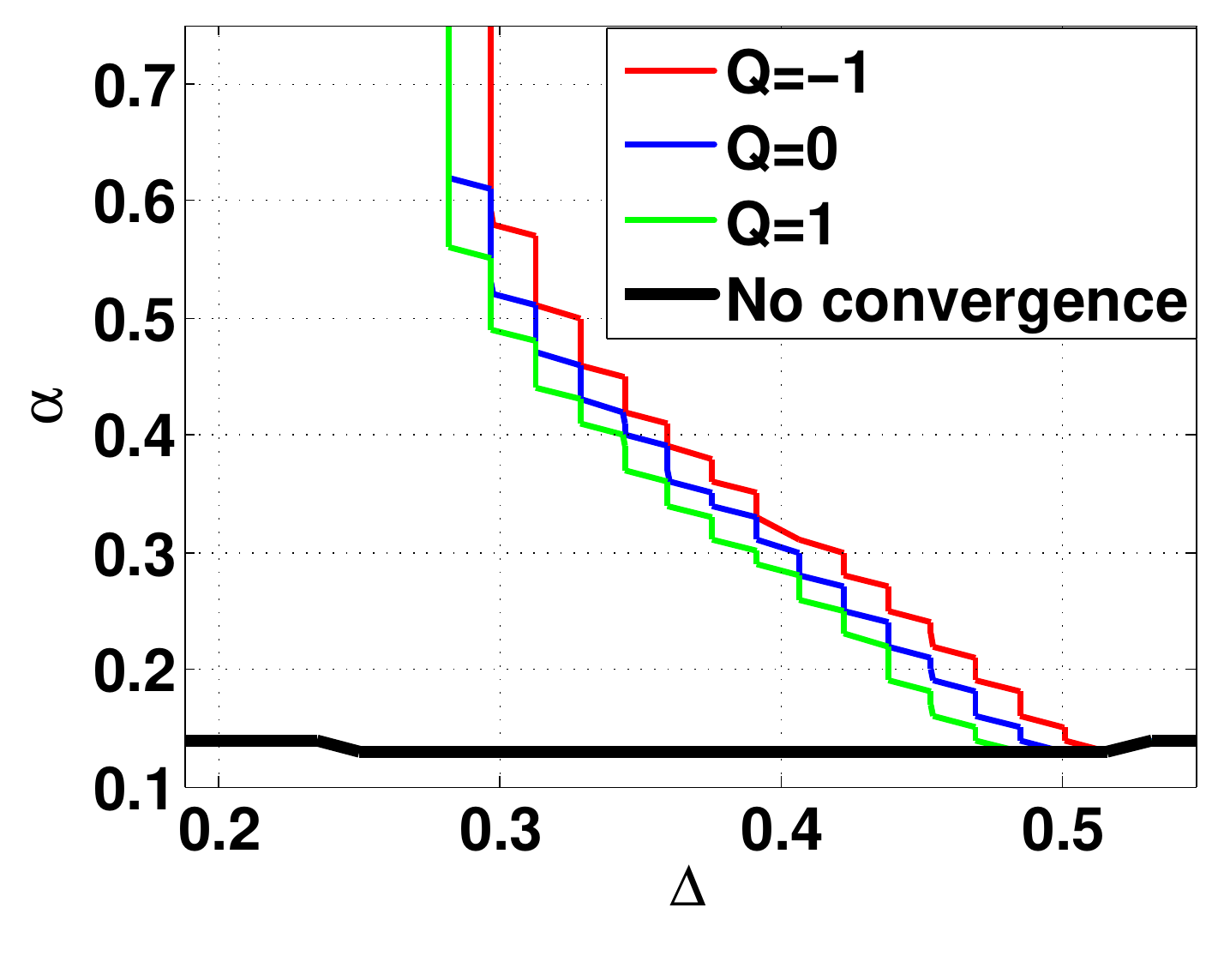}
}
\caption{(Color online) (a) The value of $\protect\alpha $ at the
bifurcation point of the spontaneous symmetry breaking of symmetric
solitons, vs. their propagation constant $k$, for different values of $Q$
and $\Delta =0.27$, as indicated in the figure. (b) The same for fixed $%
k=1.2 $ and different values of $Q$, $\Delta $ and $\protect\alpha $. The
black line bounds the region where the numerical method does not converge to
stationary solutions. Here, as well as in Figs. \protect\ref%
{fig:Twisted_Stability_map}(b) and \protect\ref{fig.Stability_2D_fundamental}
below, red lines are not shown where they overlap with blue ones, and the
latter are not shown where they overlap with green lines.}
\label{fig:Bifurcation_Split_point}
\end{figure*}

\begin{figure*}[tbp]
\centering
\subfloat[]{
            \includegraphics[width=0.49\textwidth]{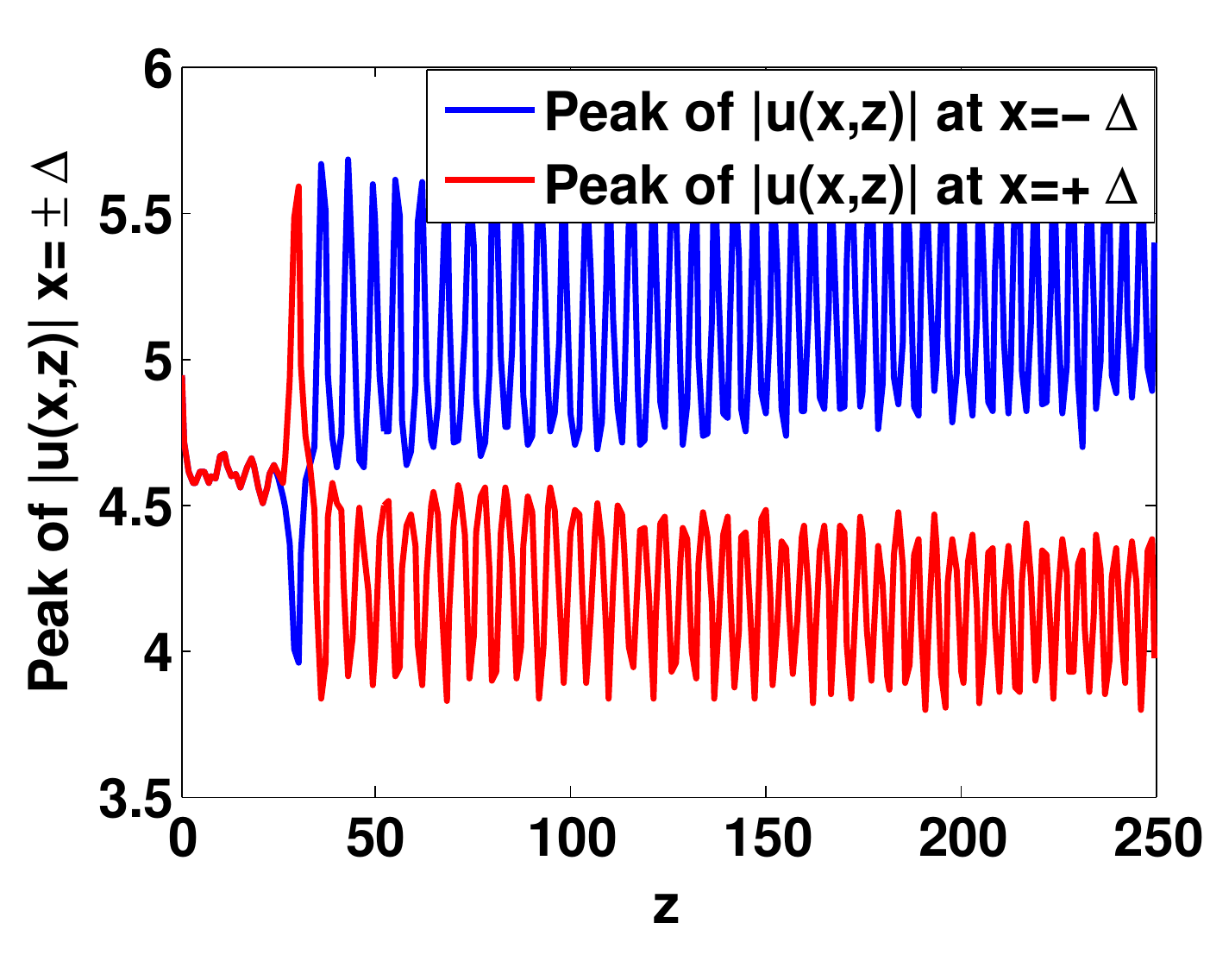}
}
\subfloat[]{
            \includegraphics[width=0.49\textwidth]{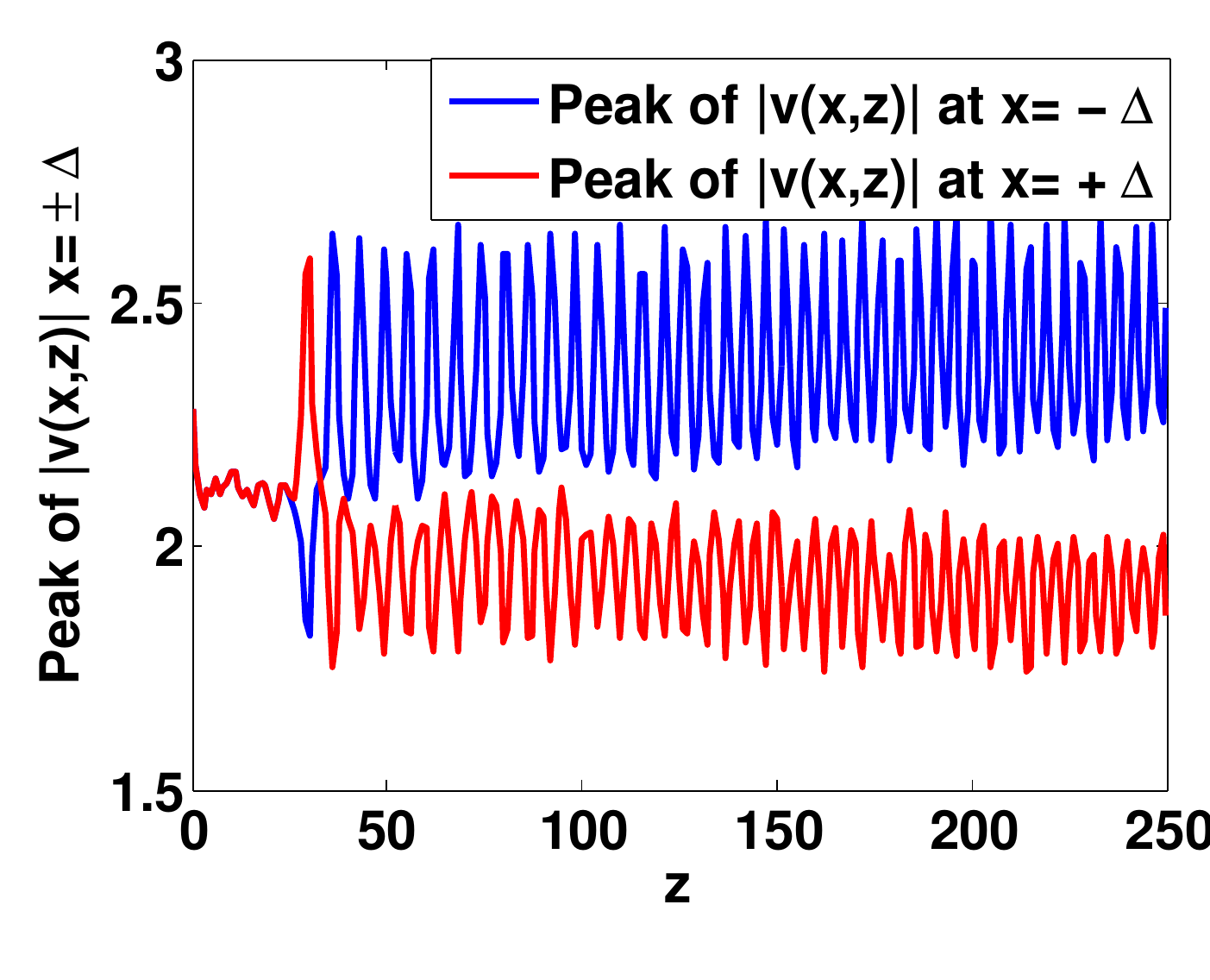}
}
\caption{(Color online) The evolution of peak powers of the two components
of an originally unstable symmetric\ soliton at $\protect\alpha =0.3,\Delta
=0.8,Q=1$, and $k=5.5$. The soliton tends to spontaneously rearrange itself
into an asymmetric state close to the co-existing asymmetric soliton.}
\label{fig:Solitons_direct_simulations}
\end{figure*}

This is a \textit{supercritical pitchfork bifurcation} \cite{Iooss}, which
destabilizes the symmetric soliton. Accordingly, the accurate analysis of
the stability demonstrates that the asymmetric solitons are always stable
when they exist, while symmetric ones are stable and unstable when they,
respectively, do not or do coexist with the asymmetric solitons. Direct
simulations [see Fig. \ref{fig:Solitons_direct_simulations}] demonstrates
that the unstable symmetric solitons spontaneously transform into asymmetric
breathers oscillating around coexisting stable asymmetric solitons.

\begin{figure*}[tbp]
\centering
\includegraphics[width=0.5\textwidth]{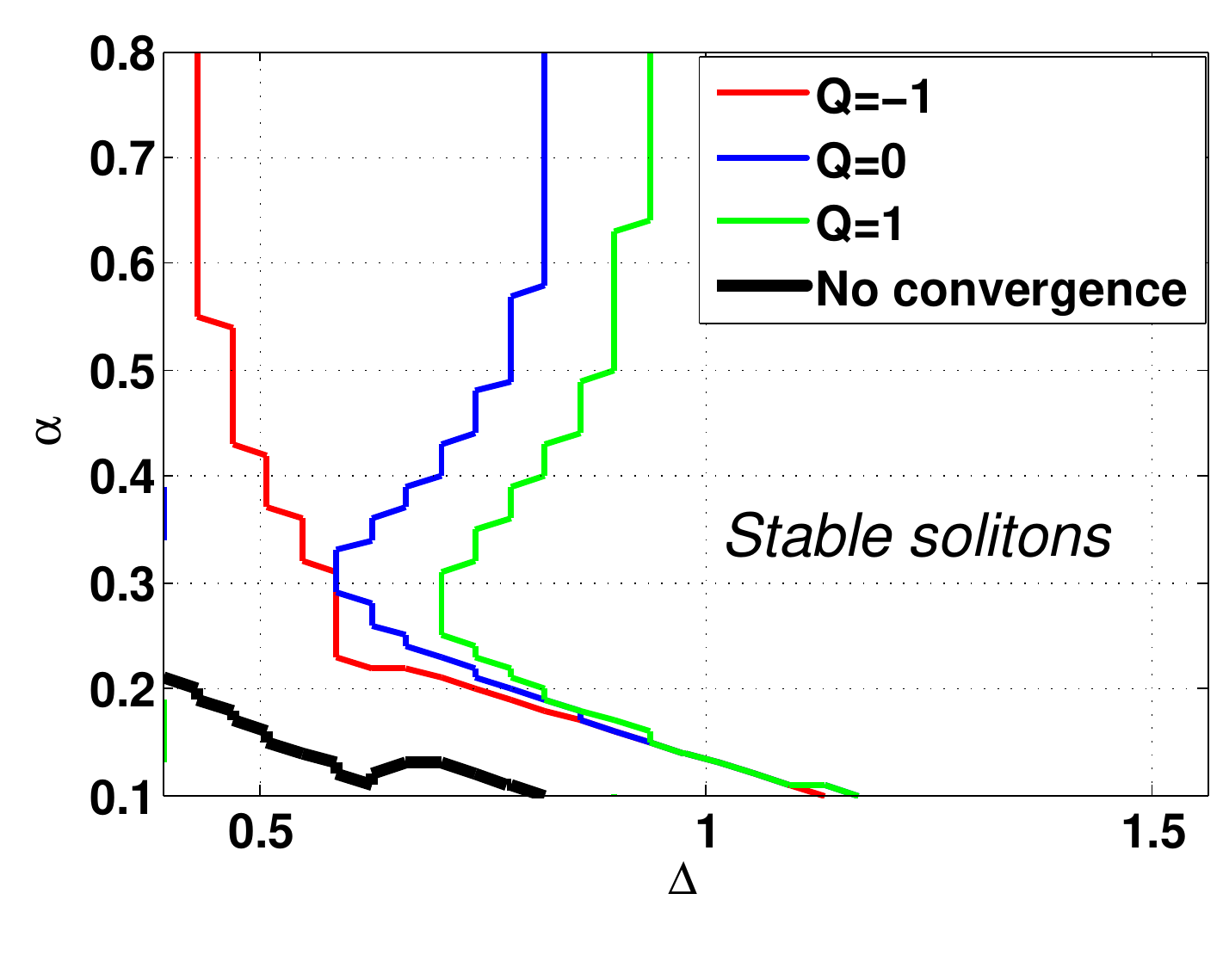}
\caption{(Color online) The stability map for twisted solitons with
different values of $Q$ and fixed wavenumber $k=1$. The solitons are stable
at values of $\Delta $ exceeding those corresponding to the displayed
boundaries. The region where the numerical method does not converge to a
stationary solution is bounded by the black line (the leftmost one).}
\label{fig:Twisted_Stability_map}
\end{figure*}

As concerns twisted solitons, they do not undergo breaking of the
antisymmetry, and are stable for sufficiently large values of distance $%
2\Delta $ between the peaks. An example of the stability map for twisted
solitons is presented in Fig. (\ref{fig:Twisted_Stability_map}), which
demonstrates that the twisted solitons are stable at $2\Delta \gtrsim 1.5$.
In direct simulations, unstable twisted solitons tend to spontaneously
transform into stable symmetric or asymmetric ones coexisting with them (not
shown here in detail).

\section{ Two-dimensional \textbf{solitons and vortices}}

\subsection{Analytical estimates}

In the 2D setting, axially symmetric solutions of stationary equations (\ref%
{phi_2d_stationary}) and (\ref{psi_2d_stationary}), with integer vorticity $%
m $ (fundamental 2D solitons correspond to $m=0$), are looked for, in polar
coordinates $\left( r,\theta \right) $, as
\begin{equation}
\varphi (r,\theta )=U(r)e^{im\theta },\psi (r,\theta )=V(r)e^{2im\theta },
\label{S}
\end{equation}%
with real amplitudes $U(r)$ and $V(r)$ satisfying equations

\begin{gather}
-kU+\frac{1}{2}\left[ \frac{1}{r}\frac{dU}{dr}+\frac{d^{2}U}{dr^{2}}-\frac{%
m^{2}}{r^{2}}U\right] +{r}^{-\alpha }UV=0,  \label{phi_2d_stationary_radial}
\\
-4kV+\frac{1}{2}\left[ \frac{1}{r}\frac{dV}{dr}+\frac{d^{2}V}{dr^{2}}-\frac{%
4m^{2}}{r^{2}}V\right] -QV+\frac{1}{2}{r}^{-\alpha }U^{2}=0,
\label{psi_2d_stationary_radial}
\end{gather}%
which can be derived from the Lagrangian [cf. its 1D counterpart (\ref{L})]:
\begin{widetext}
\begin{equation}
L=\frac{1}{2}\int_{0}^{\infty }rdr\left\{ \frac{1}{2}\left[ \left( \frac{dU}{%
dr}\right) ^{2}+\left( \frac{dV}{dr}\right) ^{2}\right] +\left[ \left( k+%
\frac{m^{2}}{2r^{2}}\right) U^{2}+\left( 4k+\frac{2m^{2}}{r^{2}}+Q\right)
V^{2}-r^{-\alpha }U^{2}V\right] \right\} .  \label{L2D}
\end{equation}
\end{widetext}The VA for the 2D solitons may be based on the following
ansatz (which implies $m\geq 0$),
\begin{equation}
U(r)=Ar^{m}\exp \left( -\rho r^{2}\right) ,V(r)=Br^{2m}\exp \left( -\gamma
r^{2}\right) ,  \label{2d_ansatz}
\end{equation}%
cf. the 1D ansatz (\ref{ansatz}). The substitution of the ansatz into Eq. (%
\ref{L2D}) yields the following effective Lagrangian:

\begin{widetext}
\begin{equation}
\begin{split}
L &=\frac{1}{4}\left\{ 4^{-m}B^{2}m(4k+Q)\gamma ^{-1-2m}\Gamma \left(
2m\right) +\frac{1}{2}4^{-m}B^{2}\gamma ^{-2m}\Gamma \left( 2+2m\right)
\right\} + \\
&\frac{1}{8}A^{2}\left\{ 2^{-m}\rho ^{-1-m}\left[ k\Gamma \left( 1+m\right)
+\rho \Gamma \left( 2+m\right) \right] -2B(\gamma +2\rho )^{-1-2m+\frac{
\alpha }{2}}\Gamma \left( 1+2m-\frac{\alpha }{2}\right) \right\} ,
\end{split}
\label{Lagrangian_2d}
\end{equation}%
\end{widetext}where $\Gamma $ is again the Gamma-function, cf. 1D Lagrangian
(\ref{lagrangian_result}). While subsequent analysis can be performed in an
explicit form for any integer vorticity $m$, eventually all the vortices
with $m\geq 1$ turn out to be unstable, only the fundamental 2D solitons
with $m=0$ having a stability area shown below in Fig. \ref%
{fig.Stability_2D_fundamental}. Therefore, explicit analytical results \ are
given here for $m=0$; nevertheless, the VA predictions for vortices have
been obtained too, and their comparison with numerical counterparts is
presented below in Fig. \ref{fig:Properties of numerically found vortex
solitons}.

The first two variational equations, $\partial L/\partial B=\partial
L/\partial \left( A^{2}\right) =0$, if applied to Lagrangian (\ref%
{Lagrangian_2d}), yield expressions for the amplitudes, cf. similar results (%
\ref{A_sqr}) and (\ref{B}) for the 1D solitons:

\begin{equation}
A=\frac{\sqrt{(4k+Q+\gamma )(\gamma +2\rho )^{2-\alpha }(k+\rho )}}{\sqrt{%
2\rho \gamma }\Gamma \left( 1-\alpha /2\right) }.  \label{A2D}
\end{equation}

\begin{equation}
B=\frac{(\gamma +2\rho )^{1-\alpha /2}(k+\rho )}{2\rho \Gamma \left(
1-\alpha /2\right) },  \label{B2D}
\end{equation}%
Taking these results into account, the two remaining variational equations, $%
\partial L/\partial \rho =\partial L/\partial \gamma =0$, amount, similar to
the 1D case [cf. Eq. (\ref{var-eqns})], to a pair of coupled quadratic
equations,

\begin{equation}
\begin{cases}
k(\gamma +\alpha \rho )-\left( 2-\alpha \right) \rho ^{2}=0, \\
\gamma (2-\alpha )\gamma +4k(\gamma -\alpha \gamma -2\rho )+Q(\gamma -\alpha
\gamma -2\rho )=0,%
\end{cases}
\label{var-eqns-2D}
\end{equation}%
which can be solved numerically.

As seen from Eqs. (\ref{A2D}) and (\ref{B2D}), these results makes sense for
$\alpha <2$ (recall in 1D the result was meaningful for $\alpha <1$).
Indeed, the average value of the $\chi ^{(2)}$ coefficient in a circle of
radius $R$ surrounding the singular point is [cf. Eq. (\ref{singular})]%
\begin{equation}
\left\langle \chi ^{(2)}\right\rangle _{\mathrm{2D}}\equiv \frac{2}{R^{2}}%
\int_{0}^{R}r^{1-\alpha }dr=\frac{2}{2-\alpha }R^{-\alpha },
\end{equation}%
hence the singular nonlinearity-modulation profile may support 2D $\chi
^{(2)}$ solitons at $\alpha <2$.

It is possible to analyze the form of the solution at $r\rightarrow 0$,
adopting an expansion similar Eq. (\ref{vicinity}), which was used in the 1D
case:%
\begin{equation}
\varphi (r)=\varphi _{0}-\varphi _{1}r^{2-\alpha }+...,~\psi (x)=\psi
_{0}-\psi _{1}r^{2-\alpha }+....  \label{expansion}
\end{equation}%
Substituting this into Eqs. (\ref{phi_2d_stationary}) and (\ref%
{psi_2d_stationary}), one can easily find%
\begin{eqnarray}
\frac{\varphi _{1}}{\varphi _{0}} &=&\frac{2\psi _{0}}{\left( 2-\alpha
\right) ^{2}},  \label{102D} \\
\psi _{1} &=&\frac{\varphi _{0}^{2}}{\left( 2-\alpha \right) ^{2}},
\label{1002D}
\end{eqnarray}%
cf. the similar 1D relations (\ref{10}) and (\ref{100}). Obviously, Eqs. (%
\ref{102D}) and (\ref{1002D}) corroborate that the 2D fundamental solitons
exist at $\alpha <2$, as at $\alpha =2$ Eq. (\ref{1002D}) dictates $\varphi
_{2}=0$, which means that the solution degenerates into a trivial one with a
flat SH field and zero FF component.

\subsection{\textbf{Numerical results for 2D solitons and vortices}}

\begin{figure*}[tbp]
\centering
\subfloat[]{
            \includegraphics[width=0.33\textwidth]{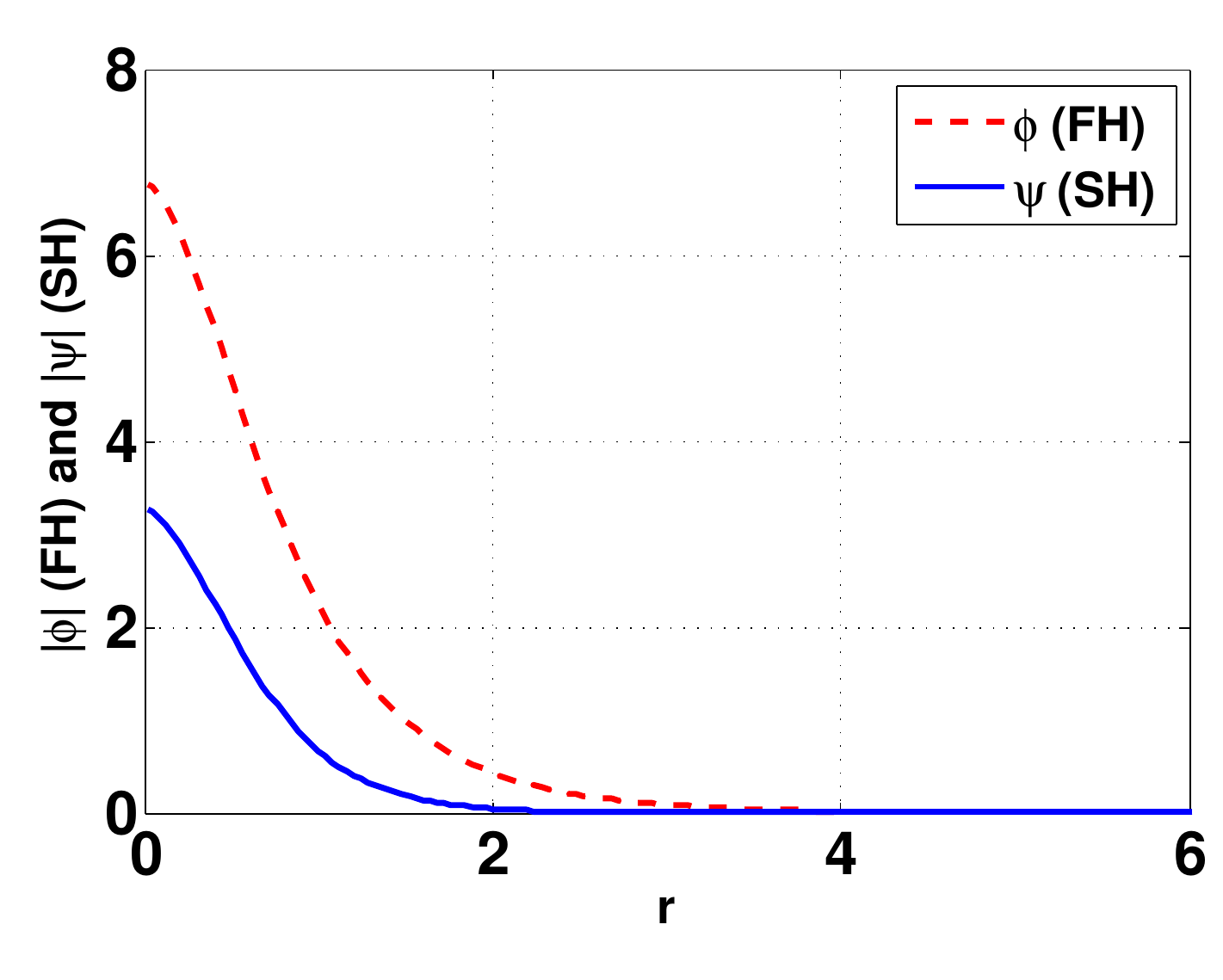}
}
\subfloat[]{
            \includegraphics[width=0.33\textwidth]{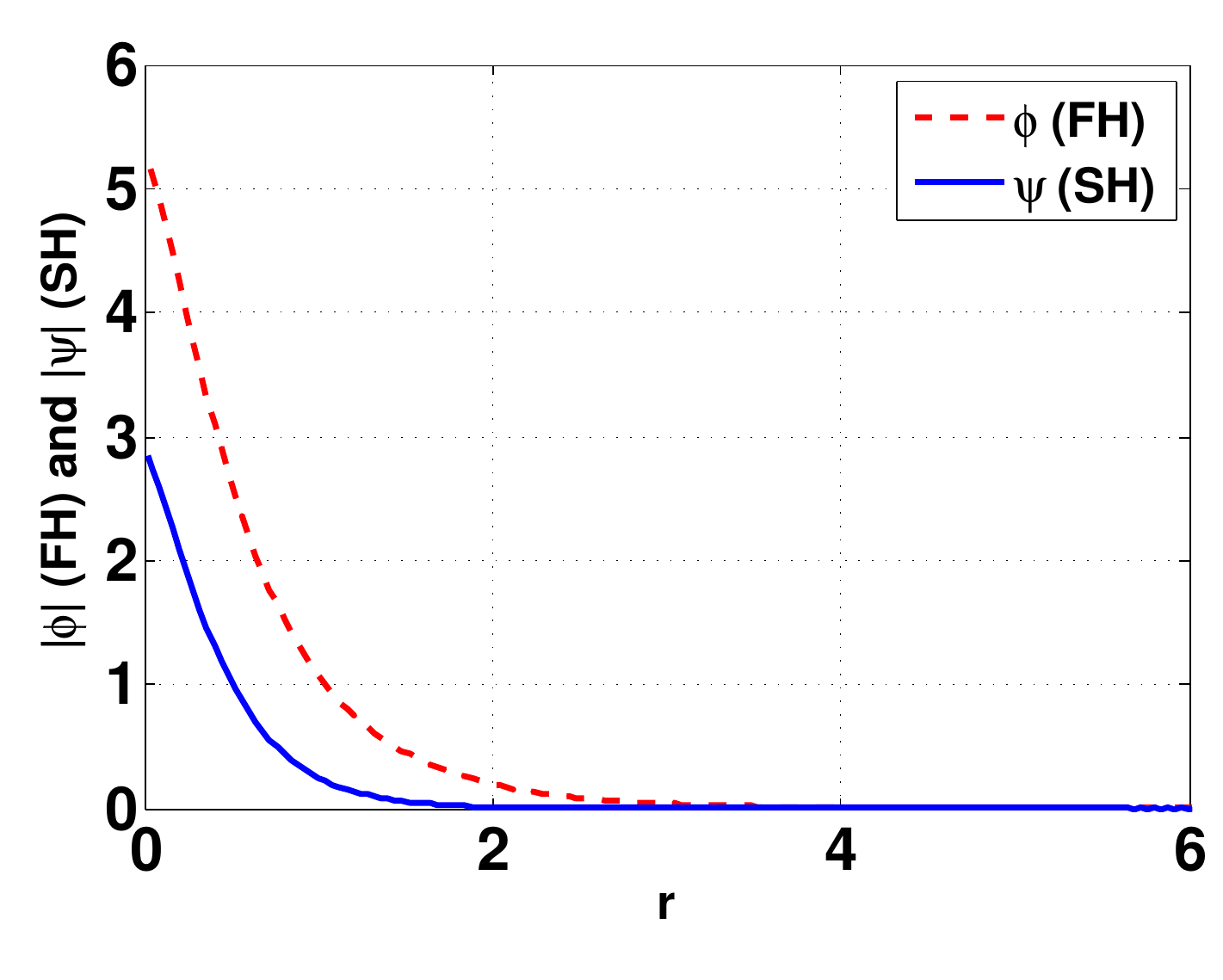}
}
\subfloat[]{
            \includegraphics[width=0.33\textwidth]{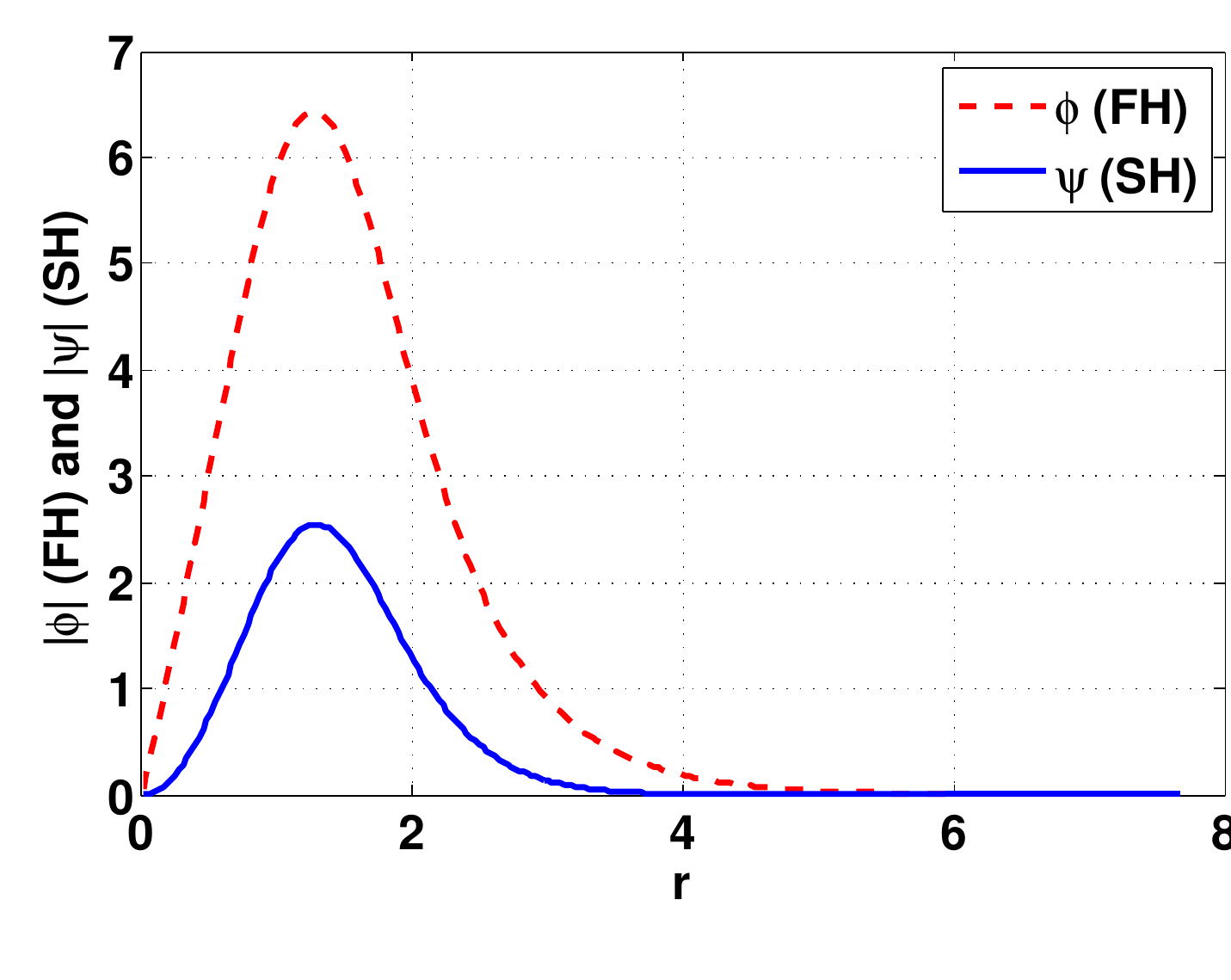}
}
\caption{(Color online) Typical radial profiles of the FF\ and SH components
of 2D solitons. (a) A stable fundamental soliton ($\protect\alpha %
=0.15,Q=1,k=1$); (b) an unstable fundamental soliton ($\protect\alpha %
=0.5,Q=1,k=1$); (c) an unstable vortex soliton with $m=1$ ($\protect\alpha %
=0.15,Q=1,k=1$).}
\label{fig:2D_Soliton_Shapes}
\end{figure*}

The Newton's method, applied to radial equations (\ref%
{phi_2d_stationary_radial}) and (\ref{psi_2d_stationary_radial}), with the
VA prediction used as the initial guess, readily generates families of
axisymmetric fundamental solitons. Solitary vortices can also be found in
the numerical form. Typical examples of radial shapes of the fundamental and
vortex ($m=1$) solitons are displayed in {Fig. {\ref{fig:2D_Soliton_Shapes}}}%
.{\ }

Properties of the 2D soliton families are summarized in Fig. \ref%
{fig:Properties of numerically found fundamental solitons}. \ Note that
dependence $P(k)$ for $Q=0$ can be found in the exact form, according to Eq.
(\ref{exact}). It is worthy to note that Fig. \ref{fig:Properties of
numerically found fundamental solitons} demonstrates that the accuracy of
the VA is actually \emph{better }for the 2D solitons than for their 1D
counterparts, cf. Fig. \ref{fig:Properties of numerically found solitons 1D}

\begin{figure*}[tbp]
\centering
\subfloat[]{
            \includegraphics[width=0.49\textwidth]{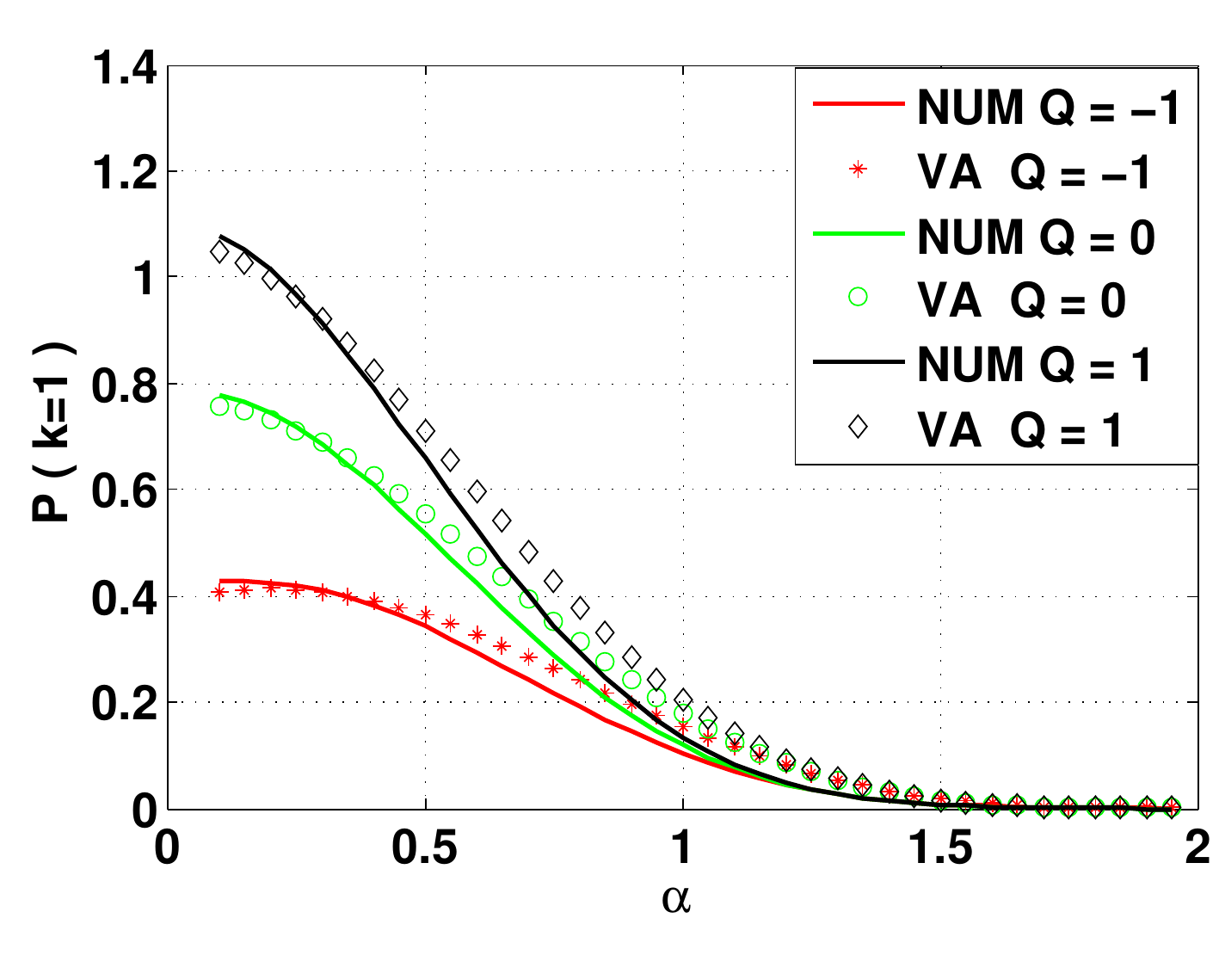}
}
\subfloat[]{
            \includegraphics[width=0.49\textwidth]{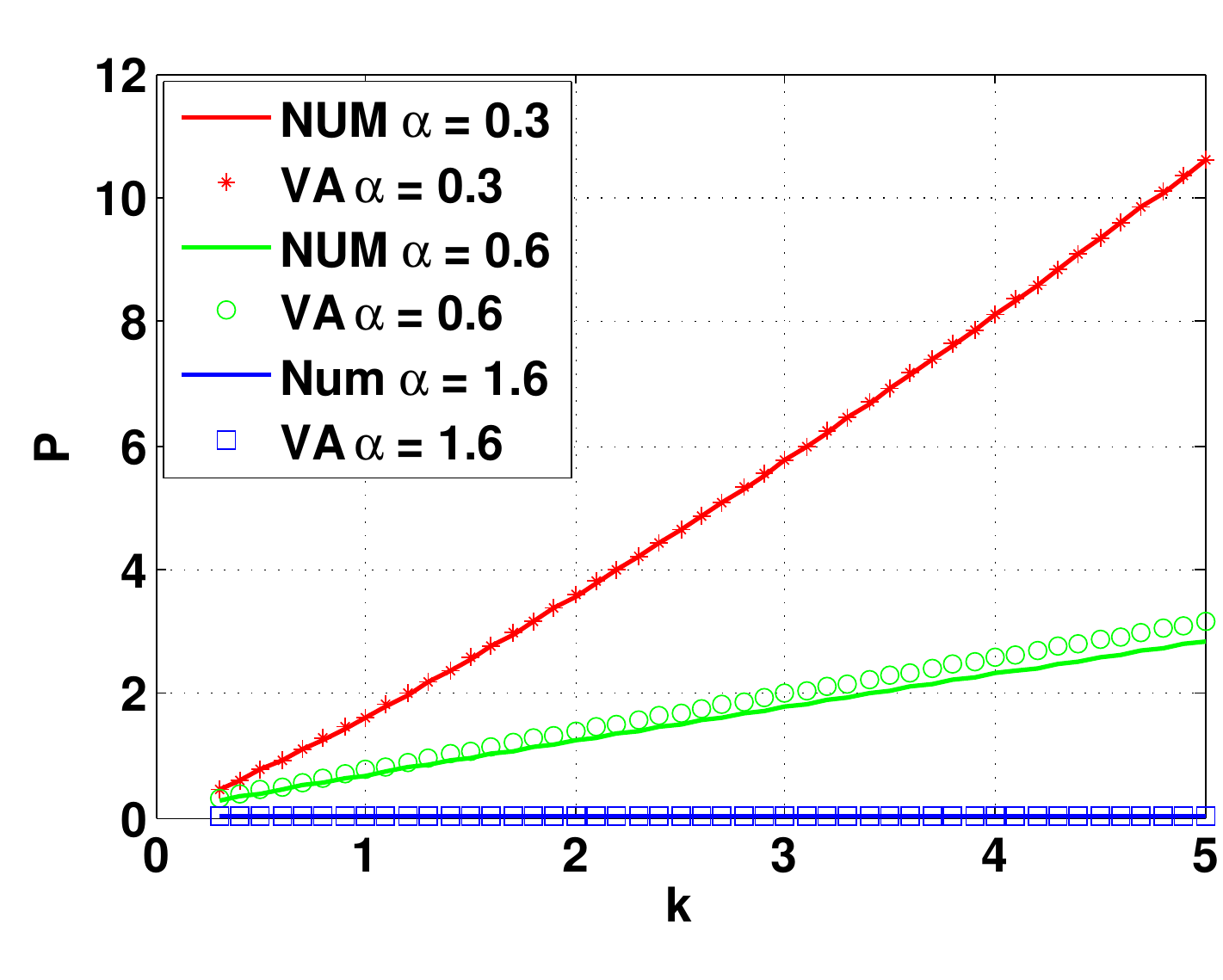}
}
\caption{(Color online) Properties of numerically found fundamental 2D
solitons, and their VA-predicted counterparts. (a) The integral power vs.
the singular-modulation exponent, $\protect\alpha $, at fixed values of
mismatch $Q$, and $k=1$. (b) The integral power vs. $k$ at several fixed
values of $\protect\alpha $ and $Q=1$.}
\label{fig:Properties of numerically found fundamental solitons}
\end{figure*}

For the sake of completeness, we display similar properties of the
vortex-soliton family with $m=1$ in Fig. \ref{fig:Properties of numerically
found vortex solitons}, although, as said above, all the vortices are
unstable. In this case, the VA still provides a reasonable overall accuracy,
although the discrepancy is larger than for the fundamental solitons.

\begin{figure*}[tbp]
\centering
\subfloat[]{
            \includegraphics[width=0.49\textwidth]{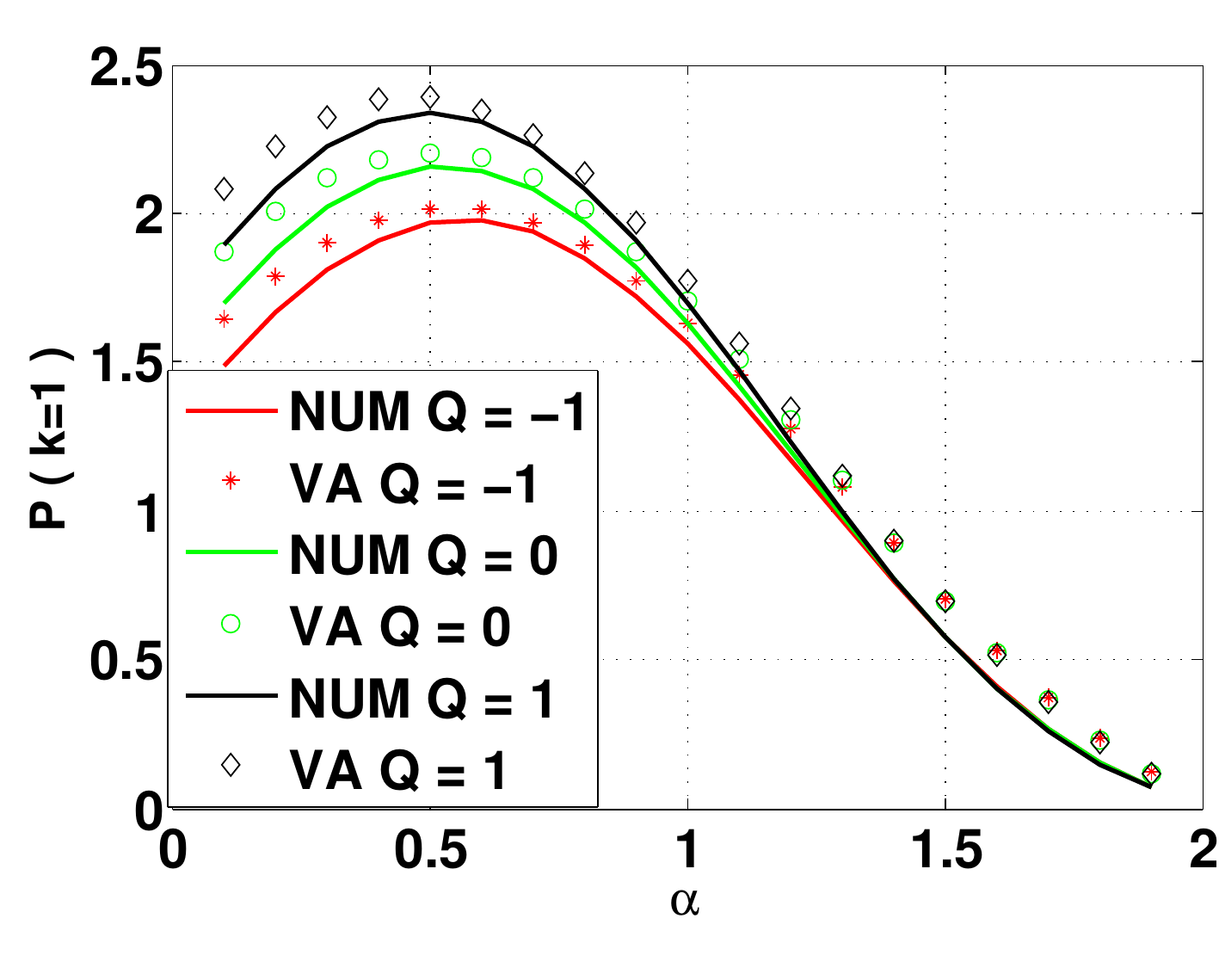}
}
\subfloat[]{
            \includegraphics[width=0.49\textwidth]{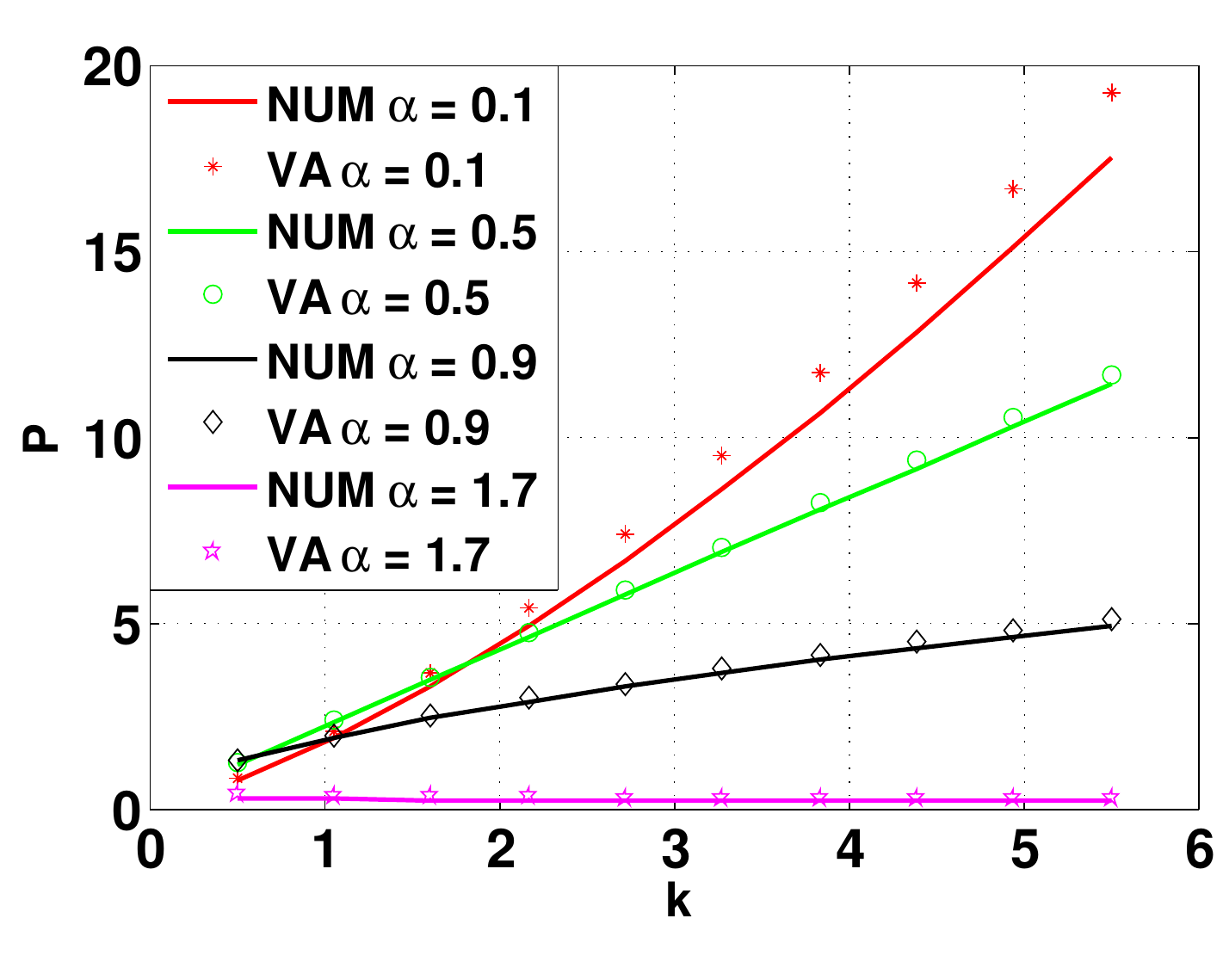}
}
\caption{(Color online) The same as in Fig. \protect\ref{fig:Properties of
numerically found fundamental solitons}, but for vortices with $m=1$. The
variational results were produced by equations derived from the effective
Lagrangian (\protect\ref{Lagrangian_2d}).}
\label{fig:Properties of numerically found vortex solitons}
\end{figure*}

The stability of the 2D solitons was analyzed by computations of eigenvalues
for modes of small perturbations, as described in Appendix \ref%
{App:AppendixB}. This analysis yields the following results. First,
stability regions for the fundamental 2D solitons are shown in Fig. \ref%
{fig.Stability_2D_fundamental}. These solitons are \emph{less stable} than
their 1D counterparts. Indeed, it was demonstrated above that the 1D
solitons have an instability region at $Q=-1$ only, see Fig. \ref%
{fig:Stability_map}, while the 2D solitons have instability regions for all
values of $Q=-1,0,1$. Moreover, it is seen in Fig. \ref%
{fig.Stability_2D_fundamental} that the stability area is slightly larger
for $Q=-1$ than for $Q=+1$. Note that, while the 2D fundamental solitons
exist up to $\alpha =2$, as shown above, the stability area is limited to $%
\alpha <0.5$.

\begin{figure*}[tbp]
\centering
\includegraphics[width=0.5\textwidth]{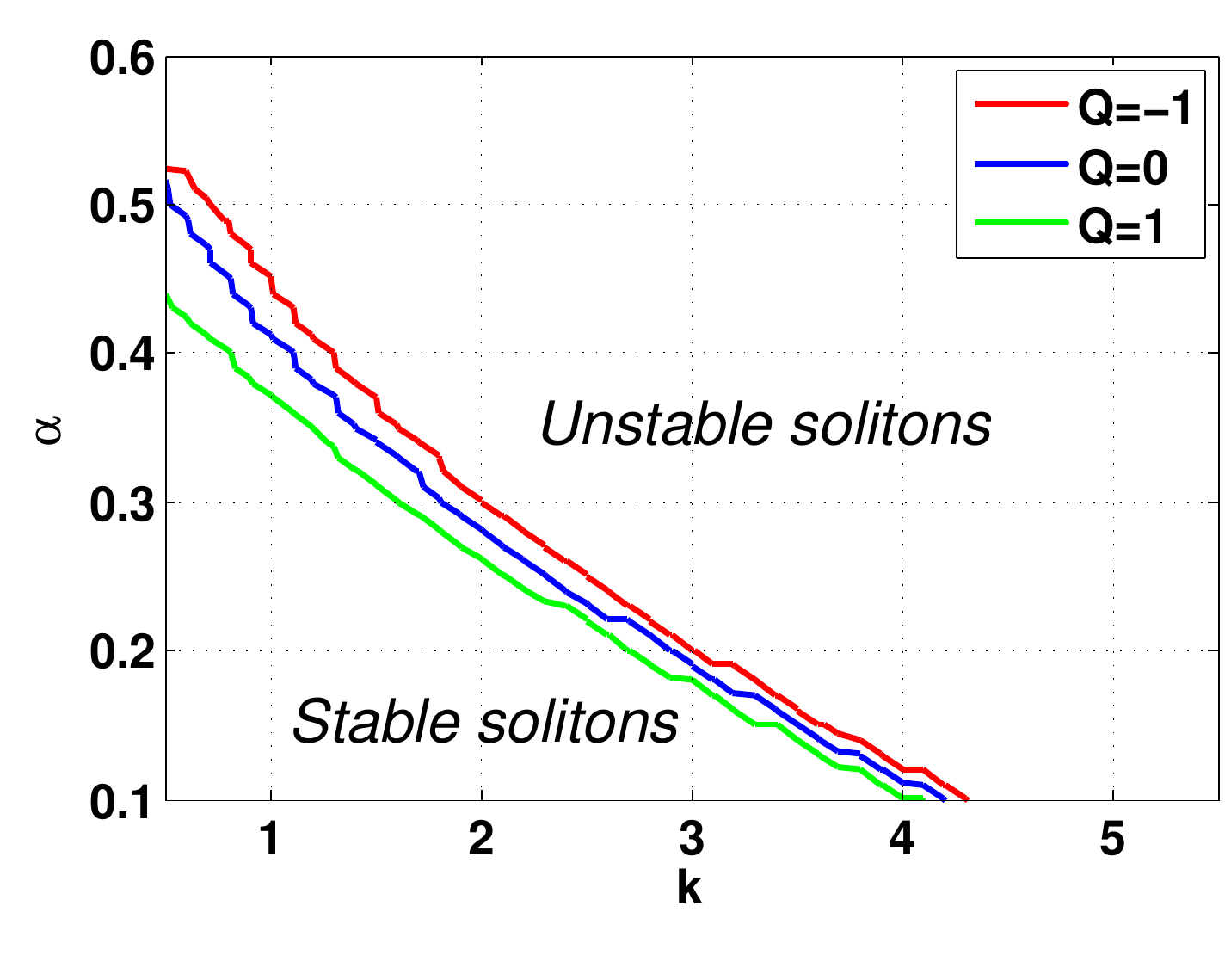}
\caption{(Color online) The stability map for 2D fundamental solitons. }
\label{fig.Stability_2D_fundamental}
\end{figure*}

Direct simulations demonstrate that the instability transforms unstable 2D
fundamental solitons into solitary breathers with a small or large amplitude
of the intrinsic oscillations, as shown in Figs. \ref%
{fig:Unstable_2d_fundamental_weak} and \ref%
{fig:Unstable_2d_fundamental_strong}, respectively.

\begin{figure*}[tbp]
\centering
\subfloat[]{
                \includegraphics[width=0.32\textwidth]{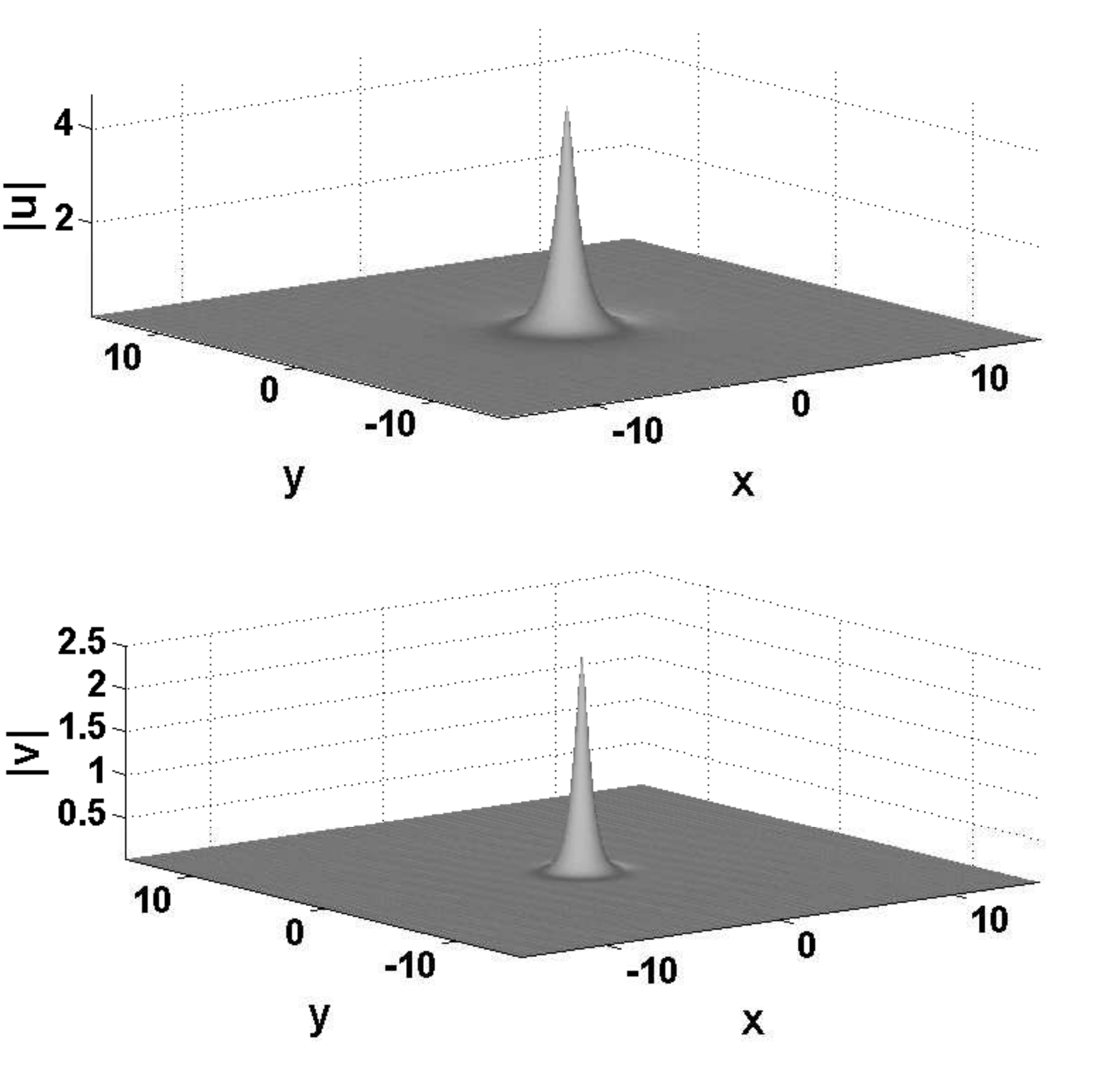}
}
\subfloat[]{
                \includegraphics[width=0.32\textwidth]{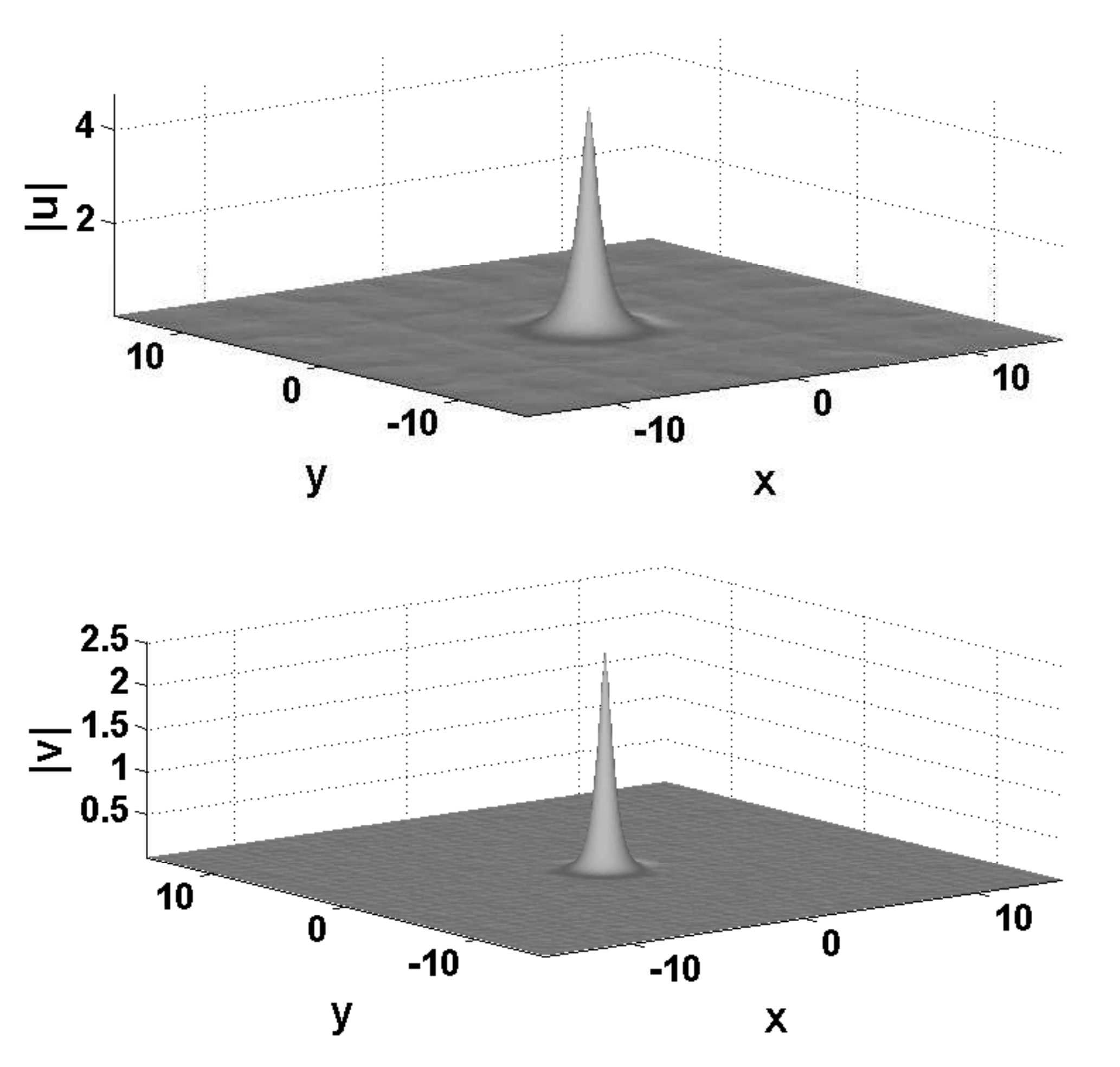}
}
\subfloat[]{
                \includegraphics[width=0.32\textwidth]{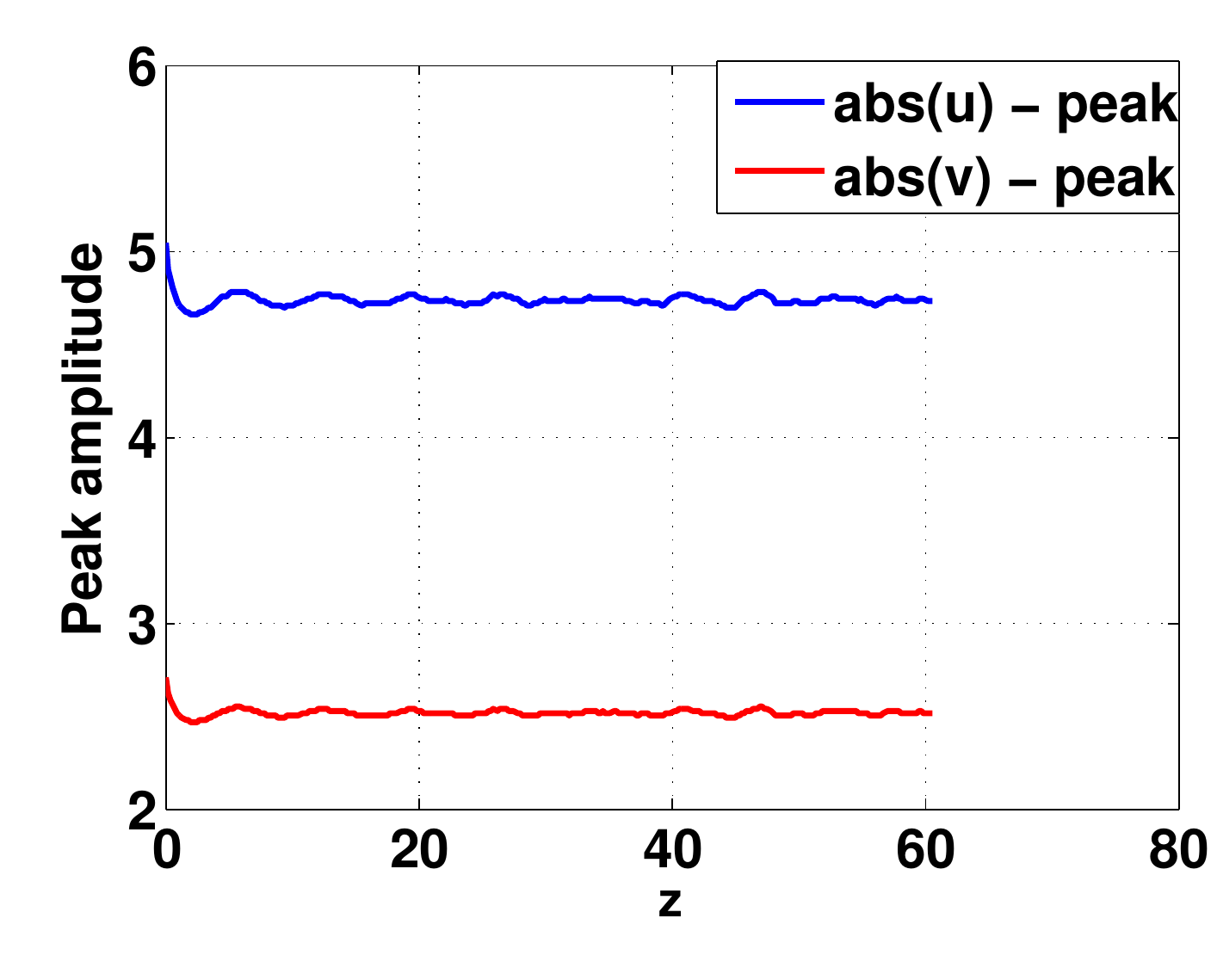}
}
\caption{An example of a breather with a small amplitude of intrinsic
oscillations, generated by the instability of a 2D fundamental soliton with $%
k=1$, $Q=1$ and $\protect\alpha =0.5$. (a,b) Shapes of absolute values of
the FF and SH fields at evolution stages corresponding to $z=4$ and $z=40$;
(c) peak values of the FH and SH fields as functions of $z$.}
\label{fig:Unstable_2d_fundamental_weak}
\end{figure*}

\begin{figure*}[tbp]
\centering
\subfloat[]{
                \includegraphics[width=0.32\textwidth]{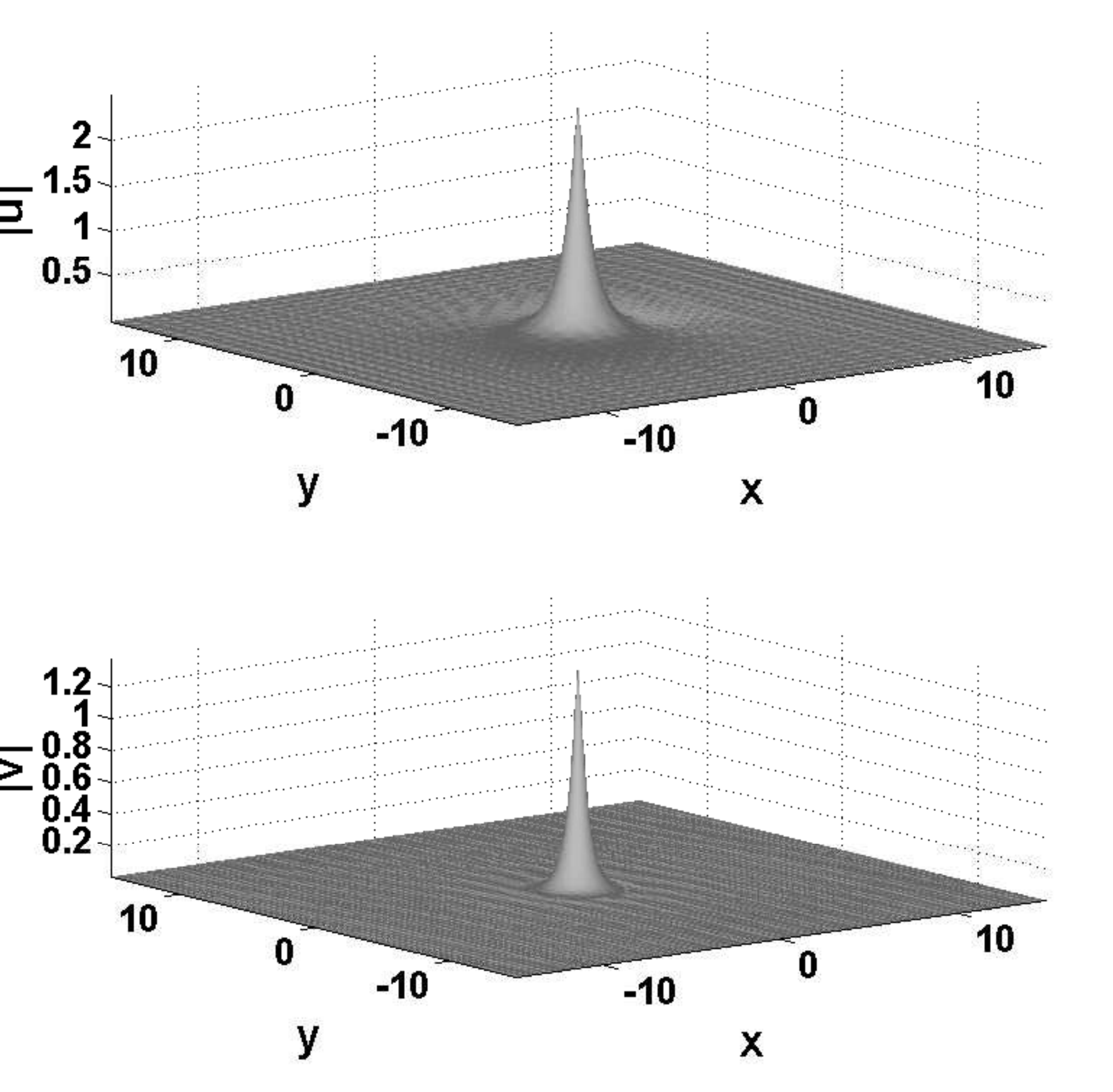}
}
\subfloat[]{
                \includegraphics[width=0.32\textwidth]{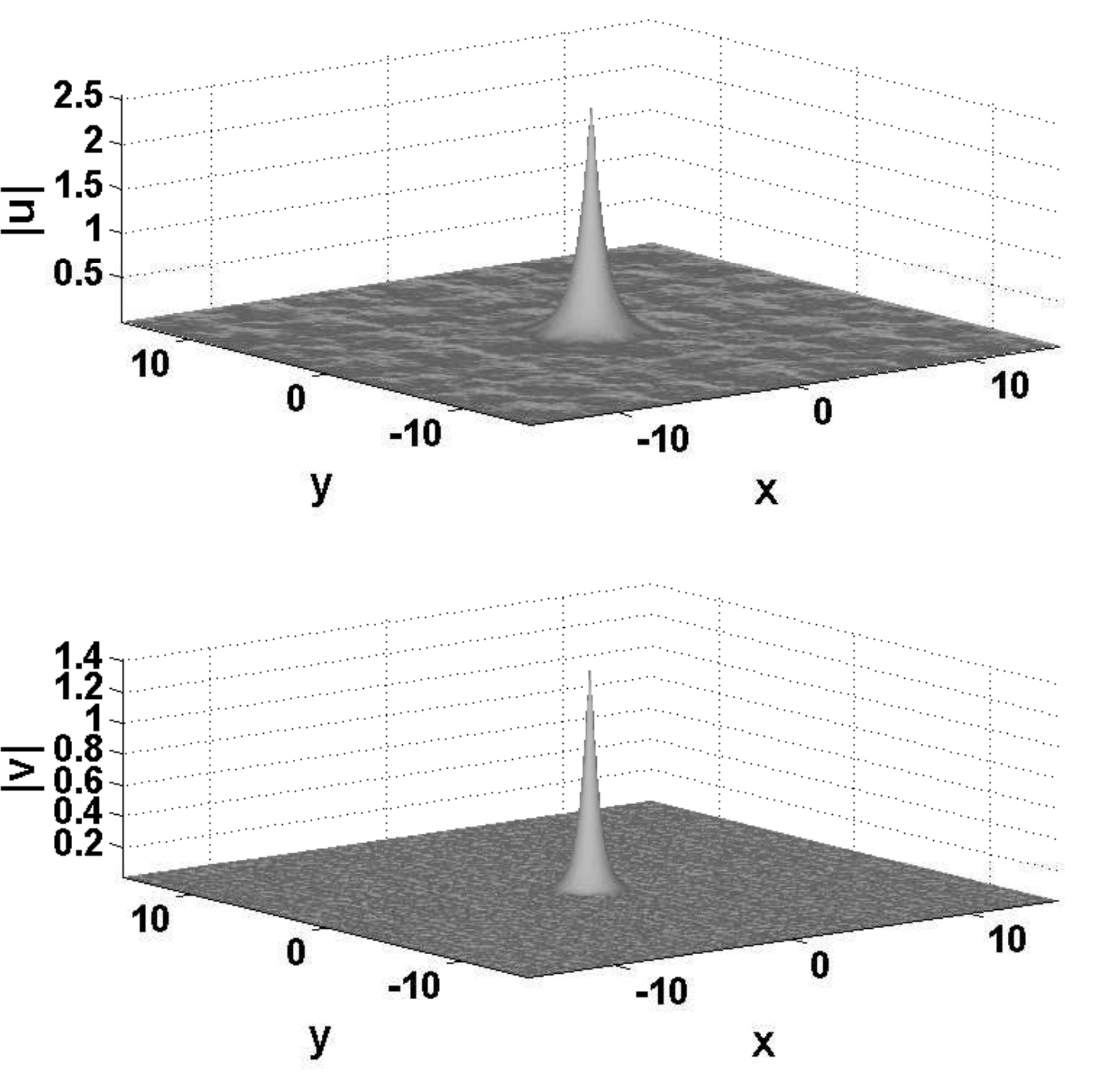}
}
\subfloat[]{
                \includegraphics[width=0.32\textwidth]{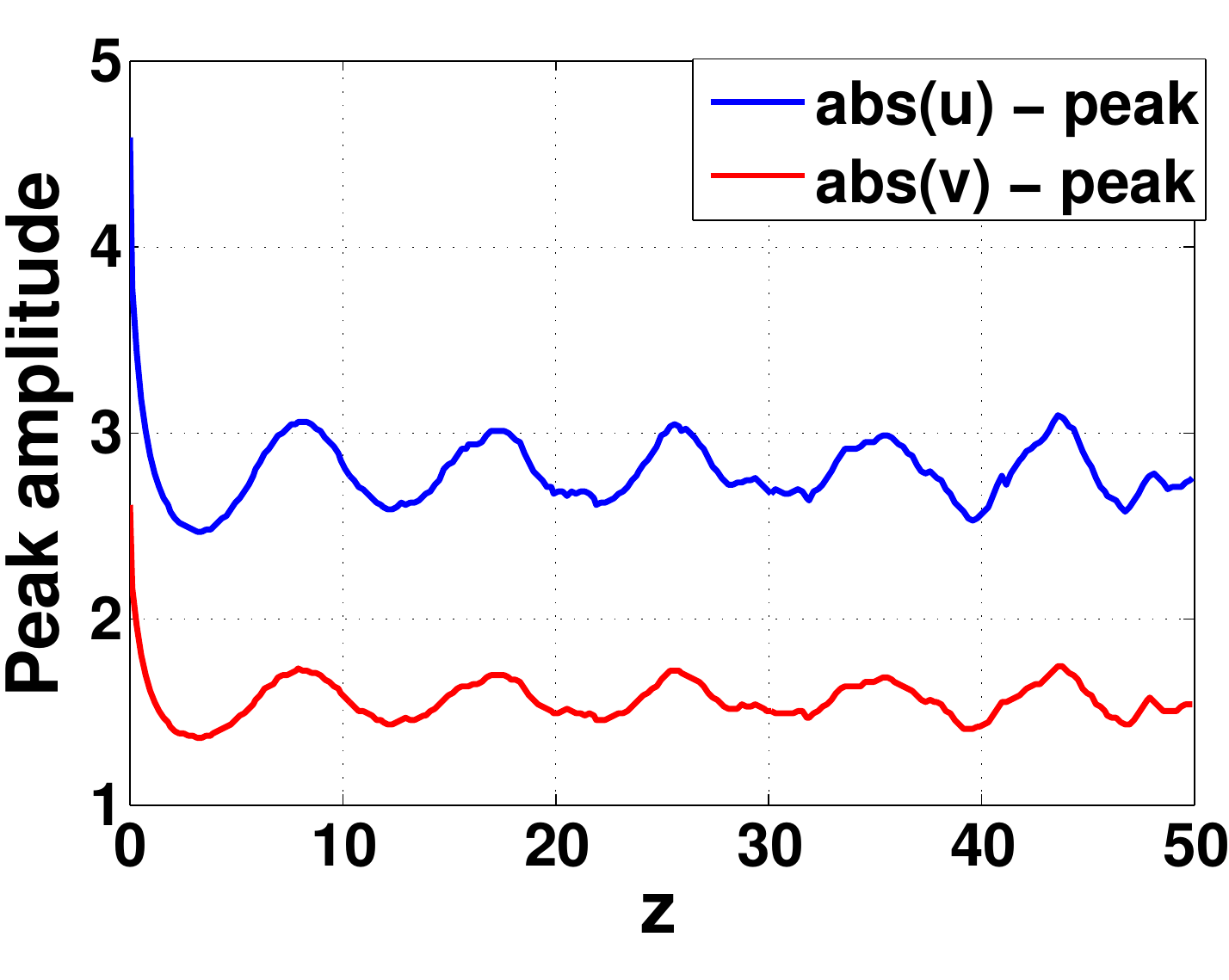}
}
\caption{The same as in Fig. \protect\ref{fig:Unstable_2d_fundamental_weak},
but for a breather with a larger amplitude of intrinsic oscillations,
generated by the instability of a fundamental soliton with $k=2$, $Q=1$ and $%
\protect\alpha =0.8$.}
\label{fig:Unstable_2d_fundamental_strong}
\end{figure*}

\begin{figure*}[tbp]
\centering
\subfloat[]{
                \includegraphics[width=0.40\textwidth]{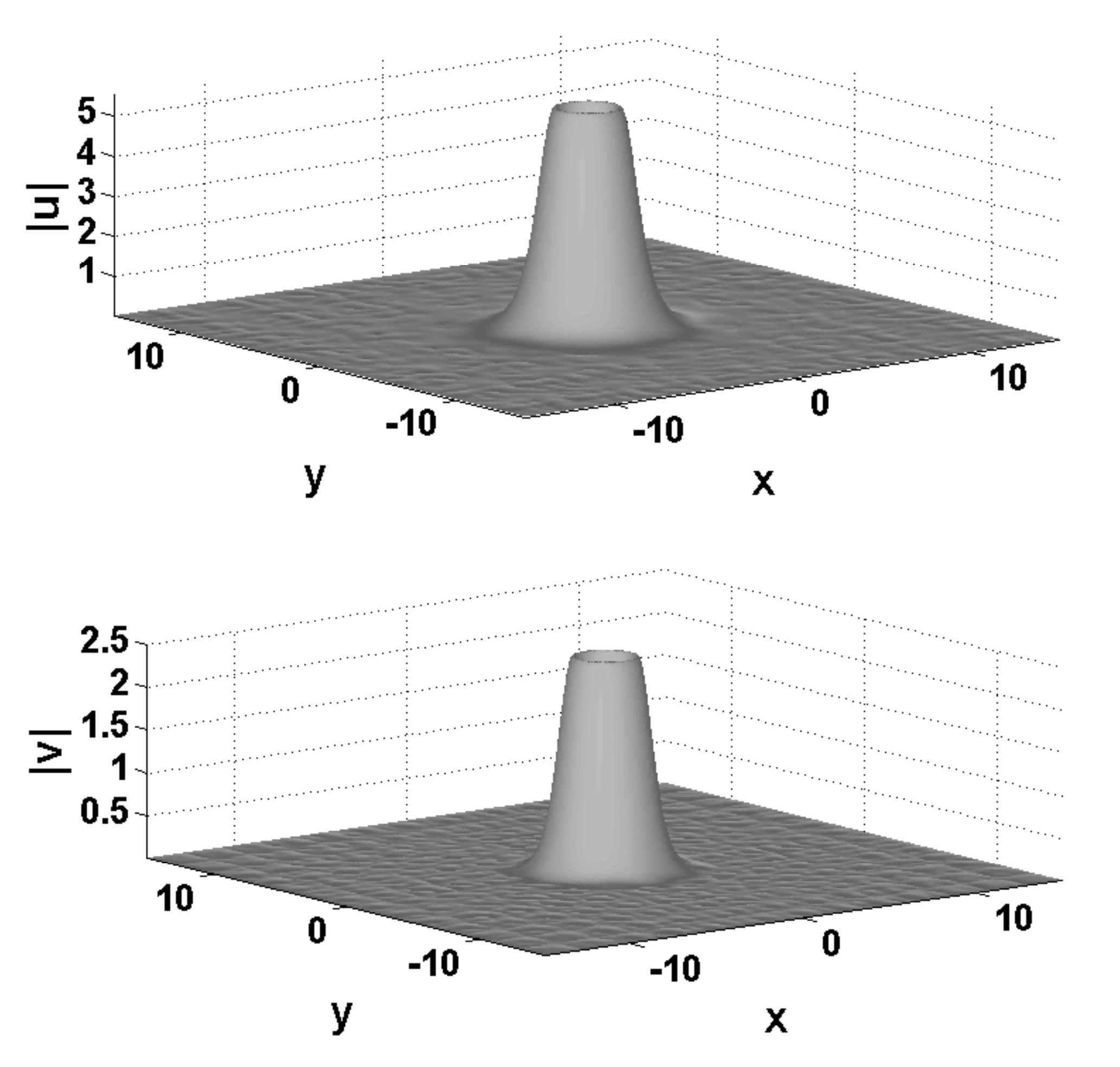}
}
\subfloat[]{
                \includegraphics[width=0.40\textwidth]{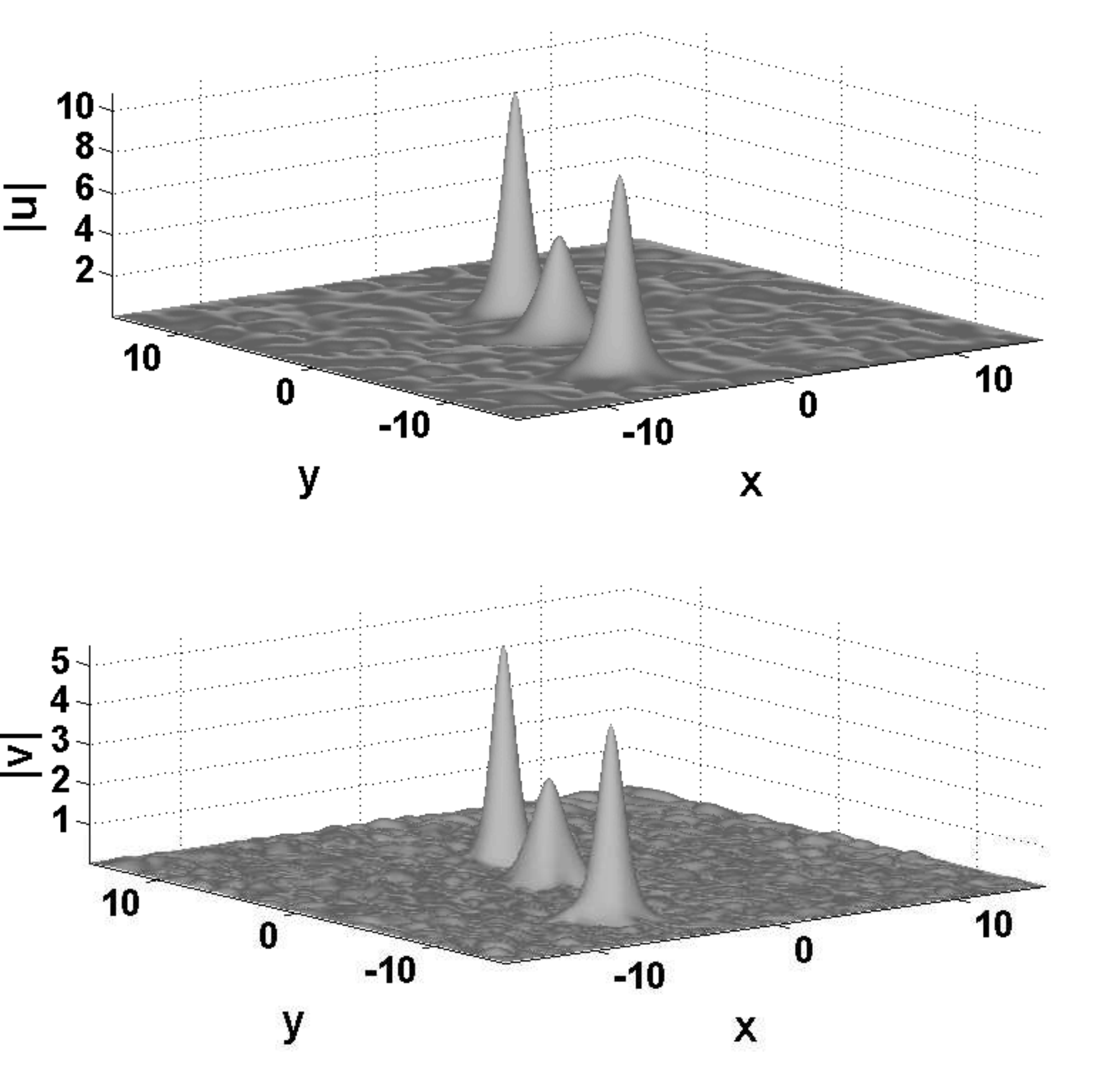}
} \newline
\subfloat[]{
                \includegraphics[width=0.40\textwidth]{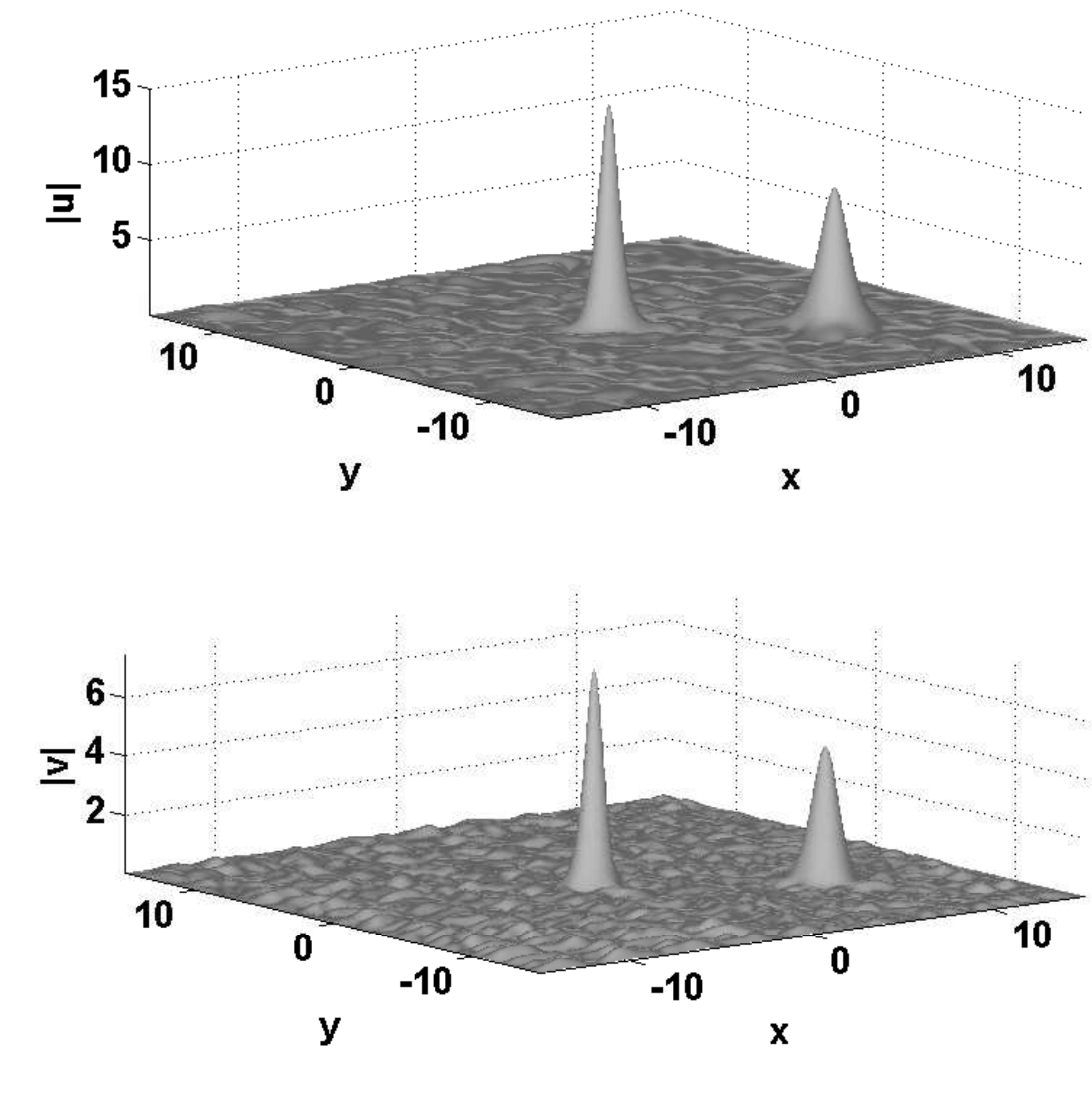}
}
\subfloat[]{
                \includegraphics[width=0.40\textwidth]{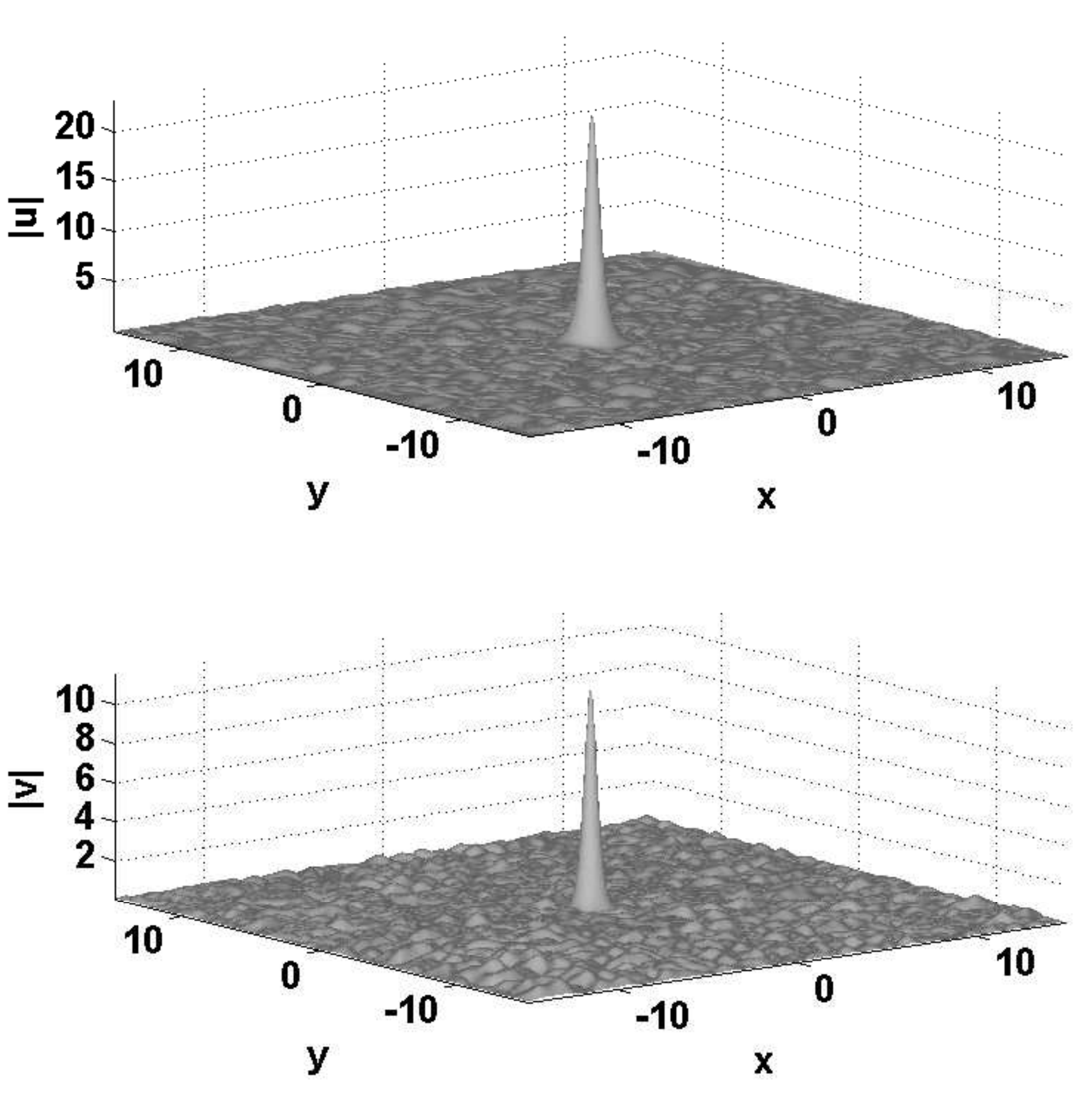}
}
\caption{An example of the instability-induced splitting of the vortex
soliton, with $m=1$, $k=1$, $Q=-1$ and $\protect\alpha =0.1$, into a set of
three fragments, which is followed by their staged merger into two
fundamental solitons, and eventually into a single one, pinned to the $%
\protect\chi ^{(2)}$ singularity. The initial instability of the vortex is
dominated by perturbation eigenmode (\protect\ref{J}) with $J=1$. Panels
(a), (b), (c) and (d) display, respectively, shapes of the absolute values
of the FF and SH fields at the evolution stages corresponding to $z=20$, $85$%
, $130$ and $230$ .}
\label{fig:Unstable_2d_examples}
\end{figure*}

On the other hand, vortex solitons are completely unstable against
eigenmodes (\ref{J}) of azimuthal perturbations, with $J=1$ and $2$ for the
vortices with $m=1$, and $J\leq 4$ for $m=2$, similar to the instability of
vortex solitons in the uniform $\chi ^{(2)}$ medium, which was studied in
detail theoretically \cite{vort-unstab} and demonstrated experimentally \cite%
{experiment}. However, direct simulations of Eqs. (\ref{u_2d}) and (\ref%
{v_2d}), which were carried out in the Cartesian coordinates, as shown in
Fig. \ref{fig:Unstable_2d_examples}, exhibit dynamics very different from
that observed in the uniform medium: the instability splits the vortex into
three fragments (two large and one smaller), which seem as fundamental
solitons. In the course of the subsequent evolution, the fragments do not
separate (as they would do in the uniform medium), but feature irregular
rotation around the $\chi ^{(2)}$ singularity (indeed, the local maximum of
the $\chi ^{(2)}$ attracts the solitons, as long as they exist). Then, two
of them collide and merge into a single soliton. Eventually, the two
remaining solitons collide twice: the first time, they bounce from each
other, but the second collision leads to their fusion into a single
fundamental soliton, which stays pinned to the $\chi ^{(2)}$ singularity.
The soliton keeps $65\%$ of the total power, being surrounded by a
conspicuous field of radiation ``debris", that carries the
entire angular momentum (\ref{M}) (we have checked that the total momentum
remains constant in the course of the evolution).

\section{\textbf{Conclusion}}

As a contribution to the currently developing studies of the dynamics of
solitons in effective nonlinear potentials, we have introduced a model of
second-harmonic-generating media with the local strength of the quadratic
nonlinearity featuring the spatial singularity, $\chi ^{(2)}\sim
|x|^{-\alpha }$ in 1D, and, similarly, $\chi ^{(2)}\sim r^{-\alpha }$ in 2D.
The spatial modulation of the nonlinearity can be implemented experimentally
with the help of the\ poling technique. We have found, using analytical and
numerical methods, that the robust fundamental solitons, pinned to the
singularity points, exist at $\alpha <1$ and $\alpha <2$, respectively, in
the 1D and 2D cases (the $\chi ^{(3)}$ counterpart of the 1D system, related
to it by the cascading limit, supports solitons in a narrower region, $%
\alpha <1/2$). The 1D solitons are stable, except for a narrow domain in the
parameter space, while the 2D fundamental solitons are stable at $\alpha
<1/2 $. A noteworthy fact is that the variational approximation provides more
accurate results for the 2D fundamental solitons than for their 1D counterparts.
In the 2D setting, vortex solitons have been constructed too. They
are unstable against splitting, but in a way different from the known
instability scenario in the uniform $\chi ^{(2)}$ medium: typically, the
unstable vortex splits into three fragments, which eventually merge into a
single soliton pinned to the central singularity. In the 1D system, we have
also studied the spontaneous symmetry breaking of solitons pinned to a pair
of singular-modulation peaks, which gives rise to nontrivial asymmetric
pinned modes via the supercritical bifurcation.

As a development of the present work, it may be interesting to study the
symmetry breaking in the 2D system with two or three symmetrically placed
singular peaks, the latter configuration being an especially interesting one
\cite{Pfau}.

\clearpage

{\appendix

\section{\textbf{Eigenmodes stability analysis of 1D model}}

\label{App:AppendixA} The stability of perturbed stationary solutions can be
evaluated by calculation of eigenvalues for modes of small perturbations.
The perturbed solutions of Eqs. (\ref{u}), (\ref{v}) are introduced as

\begin{eqnarray}
u_{p} &=&\left[ \varphi (x)+(\epsilon _{\varphi R}+i\epsilon _{\varphi I})%
\right] e^{ikz}, \\
v_{p} &=&\left[ \psi (x)+(\epsilon _{\psi R}+i\epsilon _{\psi I})\right]
e^{2ikz},
\end{eqnarray}%
where $\epsilon _{\varphi R}$, $\epsilon _{\psi R}$ and $\epsilon _{\varphi
I}$, $\epsilon _{\psi I}$ are real and imaginary parts of the perturbations.
Substituting $u_{p}$ and $v_{p}$ into Eqs. (\ref{u}), (\ref{v}), we arrive
at the system of linearized equations
\begin{widetext}
\begin{eqnarray}
i\epsilon _{\varphi _{z}}-k(\varphi _{0}+\epsilon _{\varphi })+\frac{1}{2}%
(\varphi _{0_{xx}}+\epsilon _{\varphi _{xx}})+{\left\vert x\right\vert }%
^{-\alpha }(\varphi _{0}+\epsilon _{\varphi })^{\ast }(\psi _{0}+\epsilon
_{\psi }) &=&0, \\
2i\epsilon _{\psi _{z}}-4k(\psi _{0}+\epsilon _{\psi })+\frac{1}{2}(\psi
_{0_{xx}}+\epsilon _{\psi _{xx}})-Q(\psi _{0}+\epsilon _{\psi })+\frac{1}{2}{%
\left\vert x\right\vert }^{-\alpha }(\varphi _{0}+\epsilon _{\varphi })^{2}
&=&0,
\end{eqnarray}
\end{widetext}where $\varphi _{0}$ and $\psi _{0}$ are stationary solutions
found by means of the Newton's method, $\varphi _{0}=u(x,z)_{z=0}$ and $\psi
_{0}=v(x,z)_{z=0}$, while $\epsilon _{\varphi }$ , $\epsilon _{\psi }$ are
complex perturbations. Further, the two complex coupled linearized equation
can be rewritten, in the matrix form, as a system of four real equations:
\begin{equation}
\left[
\begin{array}{c}
\epsilon _{\varphi R} \\
\epsilon _{\varphi I} \\
\epsilon _{\psi R} \\
\epsilon _{\psi I}%
\end{array}%
\right] _{z}=%
\begin{bmatrix}
0 & A & 0 & -B \\
C & 0 & B & 0 \\
0 & -D & 0 & E \\
D & 0 & -E & 0%
\end{bmatrix}%
\times \left[
\begin{array}{c}
\epsilon _{\varphi R} \\
\epsilon _{\varphi I} \\
\epsilon _{\psi R} \\
\epsilon _{\psi I}%
\end{array}%
\right] ,  \label{matrix}
\end{equation}%
with definitions
\begin{eqnarray}
A &\equiv &k-\frac{1}{2}D^{(2)}+\mathrm{diag}({\left\vert x_{i}\right\vert }%
^{-\alpha }(\psi _{0_{i}})), \\
B &\equiv &\mathrm{diag}({\left\vert x_{i}\right\vert }^{-\alpha }(\varphi
_{0_{i}})), \\
C &=&-k+\frac{1}{2}D^{(2)}+\mathrm{diag}({\left\vert x_{i}\right\vert }%
^{-\alpha }(\psi _{0_{i}})), \\
D &\equiv &\mathrm{diag}(\frac{1}{2}{\left\vert x_{i}\right\vert }^{-\alpha
}(\varphi _{0_{i}})), \\
E &\equiv &2k-\frac{1}{4}D^{(2)}+\frac{1}{2}Q.
\end{eqnarray}%
Replacing $D^{(2)}$ by the differentiation matrix and calculating
eigenvalues of resulting matrix (\ref{matrix}), the stability can be
examined.

\section{\textbf{Eigenmodes stability analysis of 2D model}}

\label{App:AppendixB} The stability analysis for the 2D model was performed
by taking perturbed solutions as
\begin{eqnarray}
u &=&\left[ U(r)+\epsilon _{u}(z,r,\theta )\right] e^{i\left( m\theta
+kz\right) }  \label{u_Perturbed_2d} \\
v &=&\left[ V(r)+\epsilon _{v}(z,r,\theta )\right] e^{2i\left( m\theta
+kz\right) }  \label{v_Perturbed_2d}
\end{eqnarray}%
and looking for perturbation eigenmodes with their own integer vorticity, $J$%
, which is independent of $m$:
\begin{equation}
\begin{cases}
\epsilon _{u}=\xi _{J}^{+}(r)e^{i(\lambda z+J\theta )}+\xi
_{J}^{-}(r)e^{-i(\lambda ^{\ast }z+J\theta )}, \\
\epsilon _{v}=\xi _{J}^{+}(r)e^{i(\lambda z+J\theta )}+\xi
_{J}^{-}(r)e^{-i(\lambda ^{\ast }z+J\theta )},%
\end{cases}
\label{J}
\end{equation}%
where $\lambda $ is the respective eigenvalue (that may be complex),
instability corresponding to $\mathrm{Im}(\lambda )\neq 0$. Numerical
solution of the eigenvalue problem, generated by the linearization of Eqs. (%
\ref{u_2d}), (\ref{v_2d}) with respect to the small perturbations, produces
the following characteristic matrix:
\begin{equation}
G=%
\begin{bmatrix}
J_{+} & A & B & 0 \\
-A & J_{-} & 0 & -B \\
\frac{B}{2} & 0 & E_{+} & 0 \\
0 & -\frac{B}{2} & 0 & E_{-}%
\end{bmatrix}%
,
\end{equation}%
where we define
\begin{widetext}
\begin{eqnarray}
J_{+} &\equiv& -k + \frac{1}{2} \left[ {\rm diag}(\frac{1}{r_{i}}) D^{(1)} + D^{(2)} - {\rm diag}(\frac{1}{r_{i}^{2}}) (J+m)^{2} \right], \nonumber \\
J_{-} &\equiv& k - \frac{1}{2} \left[ {\rm diag}(\frac{1}{r_{i}}) D^{(1)} + D^{(2)} -{\rm diag}(\frac{1}{r_{i}^{2}})(J-m)^{2} \right], \nonumber \\
E_{+} &\equiv& -2 k + \frac{1}{4} \left[ {\rm diag}(\frac{1}{r_{i}}) D^{(1)} + D^{(2)} -{\rm diag}(\frac{1}{r_{i}^{2}})(J+2m)^{2} \right] - \frac{1}{2} Q, \nonumber \\
E_{-} &\equiv& 2 k - \frac{1}{4} \left[ {\rm diag}(\frac{1}{r_{i}}) D^{(1)} + D^{(2)} -{\rm diag}(\frac{1}{r_{i}^{2}})(J-2m)^{2} \right] + \frac{1}{2} Q, \nonumber \\
A &\equiv& {\rm diag}(r_{i}^{-\alpha}V_{0_{i}}), \nonumber \\
B &\equiv& {\rm diag}(r_{i}^{-\alpha}U_{0_{i}})
\end{eqnarray}
\end{widetext}Here $D^{(1)}$ and $D^{(2)}$ are the first- and second-order
differentiation matrices. In this case the solution is unstable if there is
eigenvalue of $G$ with $\mathrm{Im}(\lambda )\neq 0$.
}

\end{document}